\documentclass[twocolumn,]{aastex701}

\newcommand{\NaID}{\ion{Na}{1} D }
\newcommand{\CaII}{\ion{Ca}{2} }

\usepackage{tikz}
\usepackage{xcolor}
\usetikzlibrary{arrows.meta}

\newcommand{\NNI}{\ion{Na}{1} D}
\newcommand{\CCII}{\ion{Ca}{2}}

\begin{document}

\title{Tracing Interstellar Gas in Quiescent Galaxies Using \NaID and \CaII H\&K Absorption}

\author[orcid=0009-0003-4285-8072]{Arian Moghni}
\affiliation{Department of Astronomy \& Astrophysics, University of California, Santa Cruz, 1156 High Street, Santa Cruz, CA 95064, USA}
\email[show]{moghni.arian@google.com}  

\author[orcid=0000-0001-6670-6370]{Timothy M. Heckman}
\affiliation{William H. Miller III Department of Physics \& Astronomy, Johns Hopkins University, Baltimore, MD 21218, USA}
\affiliation{School of Earth and Space Exploration, Arizona State University, Tempe, AZ 85281, USA}
\email{theckma1@jhu.edu}

\author[orcid=0000-0002-4430-8846]{Namrata Roy} 
\affiliation{Raman Research Institute, Sadashivanagar, Bangalore 560080, India}
\affiliation{School of Earth and Space Exploration, Arizona State University, Tempe, AZ 85281, USA}
\email{namrata@rri.res.in}   

\author[orcid=0000-0003-1809-6920]{Kyle B. Westfall}
\affiliation{University of California Observatories, University of California, Santa Cruz, 1156 High Street, Santa Cruz, CA 95064, USA}
\email{westfall@ucolick.org}

\author[orcid=0000-0001-9742-3138]{Kevin Bundy}
\affiliation{Department of Astronomy \& Astrophysics, University of California, Santa Cruz, 1156 High Street, CA 95064, USA}
\email{kbundy@ucsc.edu}

\author[orcid=0000-0001-6248-1864]{Kate H. R. Rubin}
\affiliation{Department of Astronomy, San Diego State University, San Diego, CA 92182, USA}
\affiliation{Center for Astrophysics and Space Sciences, University of California, San Diego, La Jolla, CA 92093, USA}
\email{krubin@sdsu.edu}

\begin{abstract}
For over two decades, \NaID ($\lambda\lambda 5891.58, 5897.56$ \AA) absorption has been the primary optical tracer of cool, neutral gas in galaxies. In contrast, \CaII H\&K ($\lambda\lambda 3934.78, 3969.59$ \AA), which traces both neutral and ionized gas, has been comparatively unexplored. Here, we present a comparative study of the \NaID and \CaII H\&K interstellar absorption lines using spatially resolved MaNGA IFU spectroscopy of 140 red geyser galaxies (local quiescent systems hosting low-luminosity AGN with bisymmetric ionized outflows) and a control sample of 140 quiescent galaxies. Using Gaussian fits to the \NaID and \CaII K lines, we measure gas velocity and dispersion. We find that the \CaII reservoirs are up to $\sim 6$ times larger in area, twice as radially extended, and have $\sim 1.7-1.8$x broader line widths than \NNI. Furthermore, \NaID shows a direct dependence on dust reservoir size, but the correlation vanishes for \CCII. In only $\sim 20\%$ of galaxies do the \NaID detections extend beyond dust regions, but this fraction jumps to $\sim 70\%$ for \CCII. Using these findings, we argue that dust plays a significant role in shielding \NaID from ionizing radiation, whereas \CCII, with its higher ionization potential, is not as dependent on shielding and can more easily survive beyond regions of dust. \NaID and \CaII trace distinct gas components, and studying both together provides a more complete picture of gas flows in galaxies.

\end{abstract}


\section{\textbf{Introduction}}
Gas is a fundamental component of galaxies, providing the raw ingredients for star formation and serving as a reservoir of material that can be transported, redistributed, or removed through galactic interactions, outflows, and accretion. Galaxies host gas in a wide range of phases and physical conditions, with different temperatures, densities, and ionization states. Probing material in different conditions can provide critical information about the gas cycle in galaxies. Ionized gas could be heated through shocks from active galactic nuclei (AGN) and supernovae, or photoionized by stars and AGN. Cool and warm neutral gas traces an intermediate phase between the ionized gas responding to feedback from AGN and star formation and the cooling gas that could fuel star formation and AGN activity. Lastly, cold, molecular gas traces material at even lower temperatures, with densities high enough to directly produce stars. Despite their unique qualities and distinct roles in their galaxies, these phases are not independent from one another; they each hold an important clue about the processes shaping galaxies. If we are to gain a more complete understanding of the formation and evolution of these complex systems, studying the different phases of gas and understanding their connections in a coherent way is a major step forward.

Most studies of gas in galaxies are based on the emission it produces. Studies of the gas in absorption are complementary: the surface-brightness of the optical emission lines is weighted by the square of the gas density integrated along the line-of-sight, while the absorption-line strength depends upon the gas column density (density integrated along the line-of-sight). This means that absorption-line studies are more sensitive to low-density gas. Also, because the absorption lines are seen against the background starlight, the direction of the radial flow (outward or inward) is unambiguous. These absorption lines are produced by transitions from the ground state, and basic atomic physics places nearly all these ``resonance" lines in the space-ultraviolet. There are only a few easily detectable absorption lines at optical wavelengths, most notably the \NaID doublet ($\lambda\lambda 5891.58, 5897.56$ \AA) and the \CaII H\&K doublet ($\lambda\lambda 3934.78, 3969.59$ \AA).

In the early 2000s, systematic studies of the \NaID doublet in external galaxies began and have since burgeoned. However, the \CaII doublet has been largely unexplored. There were numerous reasons for this:

\begin{enumerate}
\item The early samples of galaxies were dusty IR-selected systems, such as ultraluminous infrared galaxies (ULIRGs). In these hosts, dust attenuation at shorter wavelengths made the study of near-UV \CaII more challenging.

\item \NaID is in a relatively clean part of the spectrum without nearby strong absorption lines. On the other hand, \CaII lies in a region with many Balmer lines. In particular, the red side of the \CaII doublet (H line) at $\lambda_{rest}= 3969.6\text{ \AA}$ sits right next to the H$\epsilon$ Balmer emission line at $\lambda_{rest}=3971.2\text{ \AA}$, causing the two to be blended.

\item Lastly, early charge-coupled devices (CCDs) had a relatively poor performance at blue wavelengths \citep{1981SPIE..290...45O}. This further made the early studies of \CaII more challenging than \NNI.
\end{enumerate}

Facing these early challenges, the field began to focus on \NaID as the primary optical absorption-line tracer of gas in galaxies, while \CaII H\&K was left behind. Early studies successfully established \NaID as a robust probe of galactic-scale outflows in the most star-forming, active, and dusty galaxies (\citealt{2000ApJS..129..493H}, \citealt{2005ApJ...631L..37R}, \citealt{2005ApJS..160..115R}, \citealt{2005ApJ...632..751R}, and \citealt{2005ApJ...621..227M}). Subsequent work extended this to AGN-driven neutral winds (e.g., \citealt{2012MNRAS.426.1574D}), and the rise of the spectroscopic surveys helped show that these outflows are prevalent among massive, highly star-forming galaxies in the local Universe (e.g., \citealt{2010AJ....140..445C} and \citealt{2019A&A...622A.188C}). Recently, with the James Webb Space Telescope (JWST), studies of \NaID have been pushed into the early universe to study AGN-driven neutral gas outflows in massive galaxies (e.g., \citealt{2026arXiv260720606S} and \citealt{2024MNRAS.528.4976D}). 

With the progressively accumulating literature, \NaID has become the default tracer of cool, neutral gas. In contrast, \CaII has been left comparatively unexplored, despite some early use (\citealt{1977MNRAS.178..451B}) and occasional application in individual galaxies or small samples (e.g., \citealt{2024Natur.630...54B} and \citealt{2026A&A...705A.125L}). Recently, \citet{2025ApJ...982..108M} have used both the \CaII and \NaID absorption to characterize the cool ISM in star-forming galaxies, but comparative studies of the two absorptions features remain rare. This scarcity of \CaII studies relative to the extensive literature on \NaID underscores how much more remains to be learned from this tracer, particularly since the two lines are often treated as interchangeable probes of the same diffuse gas. 

Still, these two tracers arise from transitions in different atoms with distinct ionization states. \NaID relies on the neutral sodium atom, whose ionization potential is only 5.14 eV, much lower than hydrogen's 13.6 eV; on the other hand, the \CaII H\&K depends on the singly ionized calcium atom, with an ionization potential of 11.87 eV. These different ionization potentials suggest that the two are not necessarily interchangeable and could trace distinct gas phases in galaxies. In particular, \NaID will only exist in regions of neutral Hydrogen, while \CaII can exist in regions of neutral or ionized Hydrogen. Another potentially important difference between Na I and Ca II is that Ca is significantly more refractory than Na and is therefore subject to stronger depletion of the gas phase onto dust grains. In fact, \citet{1952ApJ...115..227R} showed that the ratio of the Ca II to Na I column densities increases as the velocity dispersion increases: shocks can destroy dust grains and liberate Ca into the gas phase. Thus, the \NaID and \CaII lines offer complementary probes of gas in galaxies.  

In this paper, we exploit this idea by comparing \NaID and \CaII H\&K in a sample of elliptical galaxies at $z\sim 0.04$. For this analysis, we focus on red geysers, which make up $\sim 6-8\%$ of the local quiescent systems observed by MaNGA. Red geysers are quiescent galaxies that show extensive ($\sim 10 \text{ kpc}$) bisymmetric features in their spatially resolved maps of ionized gas (as shown in Figure \ref{fig:DESI images and Ha EW}), with velocities far higher than, and misaligned with, the galaxy's stellar rotation (\citealt{2016Natur.533..504C}, \citealt{Roy_2018}, and \citealt{2021ApJ...913...33R}). Detailed analyses of the kinematics of these structures have shown that they are biconical winds driven from the center (\citealt{2021ApJ...913...33R} and \citealt{2019MNRAS.485.5590R}), while radio observations of the same systems point to low-luminosity AGN activity as the likely driver of these outflows (\citealt{Roy_2018} and \citealt{2021ApJ...922..230R}). \citet{2021ApJ...919..145R} were the first to study \NaID in red geysers, finding signatures of inflowing cool gas. In \citet{2026ApJ..1003...31M}, we followed up the analysis of \NaID in red geysers and provided evidence that this inflowing cool gas is fueling AGN activity in the center. As for its origins, we showed that interactions with nearby neighbors provided an efficient mechanism for importing cool gas. Here, we seek to extend that analysis to \CaII to better understand the similarities and differences of the two tracers. We also add a matched control sample of quiescent galaxies that lack the bisymmetric ionized-gas features of red geysers. By comparing \NaID and \CaII H\&K in both samples, we can test not only whether the two tracers agree within red geysers, but also whether any differences between them are specific to red geysers or exist more generally across the quiescent galaxy population.

The structure of this paper is as follows. In Section \ref{sec:Data and Sample Selection}, we describe the MaNGA survey and explain the defining features and characteristics of red geyser galaxies with additional details about our control sample. In Section \ref{sec:methods}, we outline our methodology for stellar continuum removal, Gaussian fitting, and error estimation. In Section \ref{sec:results}, we present our results on comparisons between the kinematics and prevalence of \NaID and \CaII in red geysers and the control sample and their connection to dust. In Section \ref{section:discussion}, we discuss the insights revealed from our results regarding these tracers. Lastly, in Section \ref{sec: Summary}, we summarize our findings.

\section{\textbf{Data and Sample Selection}} \label{sec:Data and Sample Selection}
\subsection{MaNGA Survey}
For this work, we use integral field spectroscopy from the Sloan Digital Sky Survey IV(SDSS-IV) Mapping Nearby Galaxies at Apache Point Observatory (MaNGA; \citet{Albareti_2017}; \citet{2017AJ....154...28B}; \citet{2015ApJ...798....7B}; \citet{2015AJ....149...77D}; \citet{2015AJ....150...19L}; \citet{Yan_2016}). MaNGA includes observations of 10,010 nearby ($z\sim 0.03$) galaxies using the Sloan Digital Sky Survey 2.5 m telescope \citep{Gunn_2006} and the BOSS spectrographs \citep{Smee_2013}. The MaNGA data provide spatially resolved spectroscopy from the near-ultraviolet to near-infrared wavelength range ($3600$--$10,000\text{ \AA}$), with a spectral resolution increasing from $R\approx1400$ at $4000\text{ \AA}$ to $R\approx2600$ at $9000\text{ \AA}$. The survey achieves an effective spatial resolution of $\sim2.5''$ (FWHM), with each data cube coming in $0.5''\times0.5''$ spatial pixels (spaxels), each containing a full spectrum across the wavelength range. For the sample of galaxies in our study, the typical projected physical area of a spaxel is $\sim 0.1\text{ kpc}^2$.

Throughout this paper, we use the same MaNGA data as \citet{2026ApJ..1003...31M} (Paper I). Additional details regarding the survey design, observations, and the data can be found in the references above and in Paper I.

\subsection{Red Geyser Sample}
\label{subsec:red geyser sample}

In this work, we trace the cool and warm gas in the interstellar medium in nearby quiescent galaxies. Our sample consists of red geysers and a matched control sample of quiescent galaxies. 

We analyze the same sample of 140 red geyser galaxies as previously presented by \citet{Roy_2018} and \citet{2021ApJ...922..230R}. Red geysers are a class of nearby ($z\sim 0.04$) quiescent galaxies that host low-luminosity active galactic nuclei (AGN) and comprise about $6-8\%$ of the local quiescent galaxy population observed in the MaNGA survey. They were first identified through visual inspection of MaNGA observations based on their unique ionized gas morphology and kinematics \citep{2016Natur.533..504C}.

Observationally, the characterizing feature of a red geyser is its bi-symmetric features in the spatially resolved equivalent width (EW) maps of strong emission lines, such as H$\alpha$, [NII], and [OIII]$\lambda5007$ (Figure \ref{fig:DESI images and Ha EW}). These are galaxy-scale ionized outflows extending over $\sim 10 \text{ kpc}$ and are accompanied by high gas velocities that reach $\pm 300 \text{ km s}^{-1}$ \citep{2021ApJ...913...33R}. An additional feature of this ionized gas in red geysers is that they have large velocity dispersions, usually around $220-250 \text{ km s}^{-1}$. These properties hint at AGN feedback in action, and studies in the radio (\citealt{Roy_2018}, \citealt{2021ApJ...922..230R}) and the modeling of these outflows \citep{2021ApJ...913...33R} have confirmed that red geysers host low-luminosity AGN, which are likely driving their observed large-scale winds.

Detailed descriptions of the sample selection and physical properties of red geysers are provided by \citet{Roy_2018} and \citet{2021ApJ...919..145R}.

\subsection{Control Sample}
\label{subsec:control sample}
A large component of our analysis is dedicated to comparing the red geyser galaxies to a control sample of systems that were first constructed in \citet{Roy_2018} and previously used in Paper I.

Similar to red geysers, our control sample of galaxies is made up of red and quiescent systems with NUV-r$>$
5. However, unlike red geysers, these systems do not exhibit the AGN-driven bi-symmetric features that define red geysers. These control galaxies are further matched one-to-one to the individual red geysers using several criteria:
\begin{enumerate}
    \item $|\log(M_{*,~red~geyser}/M_{*,~control})|<0.2\text{ dex}$,
    \item $|z_{red~geyser}$-$z_{control}|<0.01$, and
    \item $|b/a_{red~geyser} - b/a_{control}| <0.1$.
\end{enumerate}

The stellar mass and redshift criteria ensure that we are matching red geysers in mass and redshift, while the third criterion matches axis ratios to avoid potentially comparing dust-reddened edge-on galaxies with the relatively face-on red geyser galaxies.

Here, $M_*$ is the stellar mass, $z$ is the redshift, and $b/a$ is the axis ratio from the NASA-Sloan Atlas (NSA) catalog. 

\begin{figure}[t]
    \centering
    \includegraphics[width=0.48\textwidth]{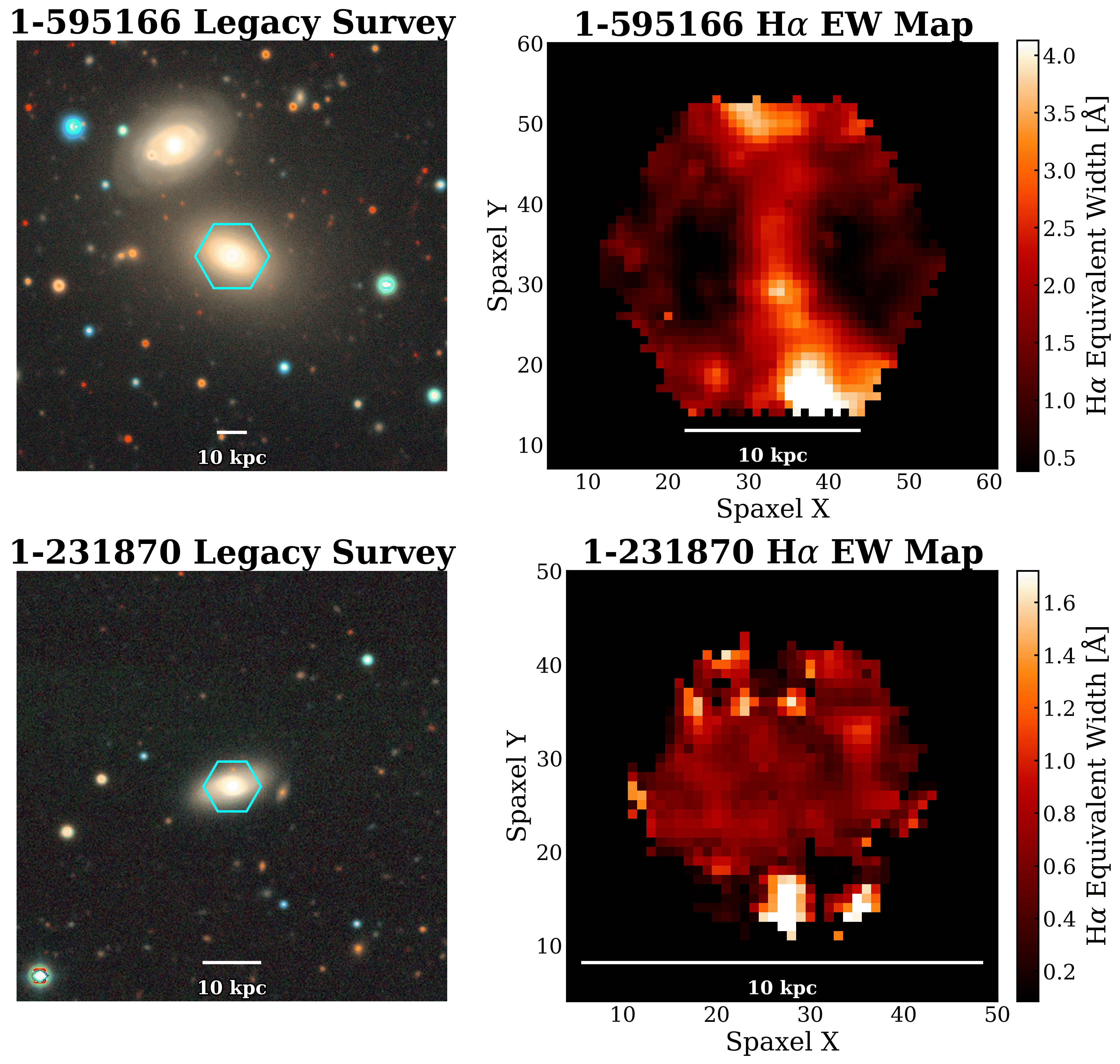} 
    \caption{DESI Legacy Survey images (left) and zoomed-in MaNGA H$\alpha$ equivalent width (EW) maps (right) for a red geyser galaxy with MaNGA ID 1-595166 at the top and a control galaxy with MaNGA ID 1-231870 at the bottom. Each Legacy Survey cutout shows the large-scale morphology and environment. The H$\alpha$ EW maps focus on the MaNGA footprint, outlined by the cyan hexagon, and highlight ionized gas and the characteristic bi-symmetric jet-like features that define red geysers.}

    \label{fig:DESI images and Ha EW}
\end{figure}

\begin{figure}[t]
    \centering
    \includegraphics[width=0.48\textwidth]{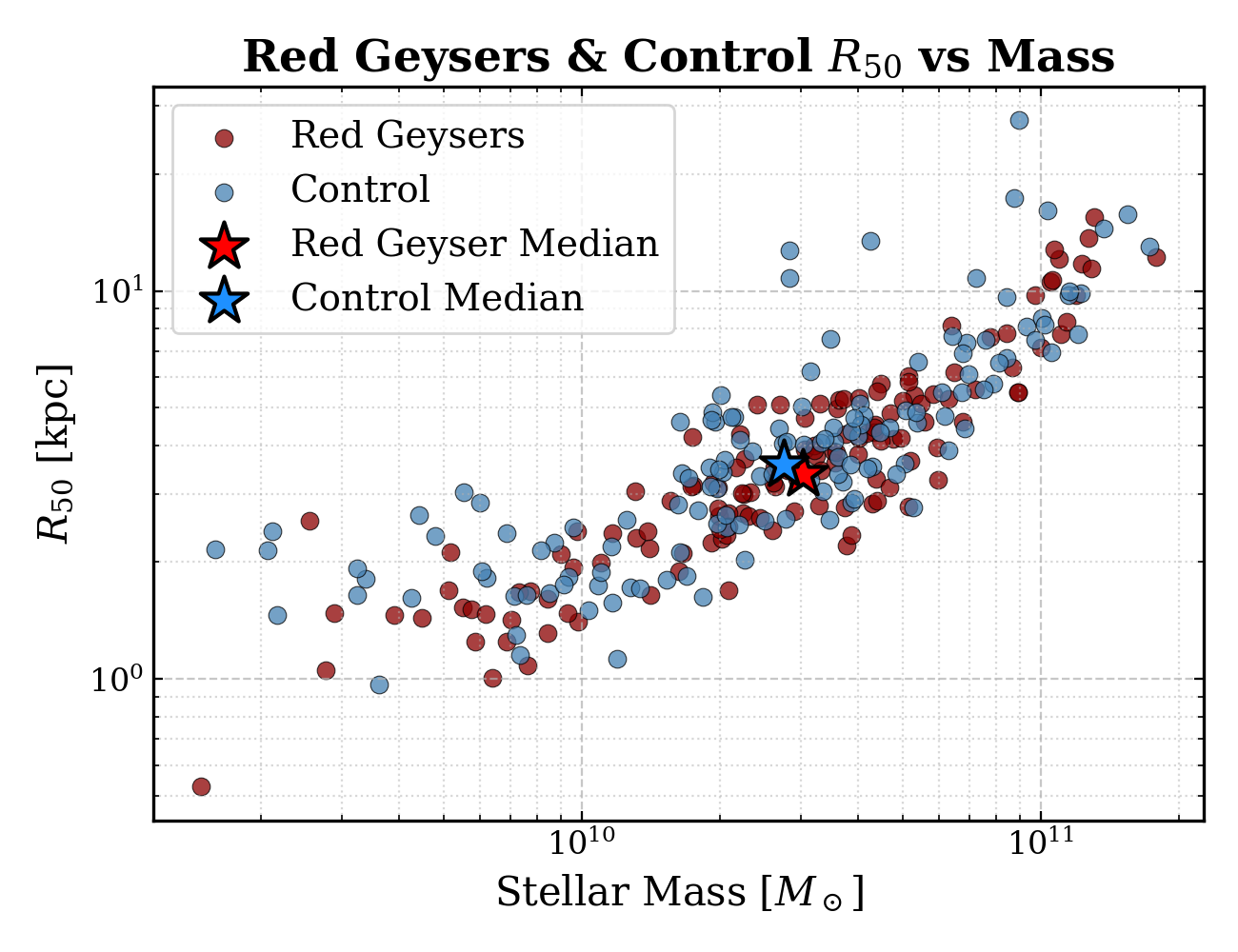} 
    \caption{Half-light radius $R_{50}$ vs. stellar mass for the full sample of red geyser galaxies in red (140 in total) compared to the control sample of galaxies in blue (also 140 in total). The red star plotted toward the middle represents a characteristic red geyser with mass $\log_{10}M_*/M_\odot =10.5$ and $R_{50}=3.4\text{ kpc}$. On the other hand, a characteristic control galaxy has a mass $\log_{10}M_*/M_\odot =10.4$ and $R_{50}=3.6\text{ kpc}$.}

    \label{fig:R50 vs stellar mass}
\end{figure}

Figure \ref{fig:DESI images and Ha EW} shows DESI \citep{2019AJ....157..168D} images of a red geyser along with a control galaxy, in addition to their MaNGA maps of H$\alpha$ EW. While the two samples' images show roughly similar elliptical and quiescent systems, the ionized gas maps for the control sample are missing the bi-symmetric features observed in red geysers. In Figure \ref{fig:R50 vs stellar mass}, the half-light radius and stellar mass of red geysers are compared with the control galaxies. The two samples show similar global properties, both having a median stellar mass of $\log_{10}(\frac{M_*}{M_{\odot}})\sim10.45$ and a median $R_{50}$ of $\sim 3.5 \text{ kpc}.$

\section{\textbf{Methods}}
\label{sec:methods}

\subsection{Removing Stellar Absorption}
\label{subsection: removing stellar absorption}
The \NaID and \CaII H\&K absorption features that we seek to study not only arise from the cool and warm gas in the interstellar medium, but they could also arise from the photospheres of stars in their host galaxies. Since we are only interested in studying this ISM gas, we must remove the stellar component. This procedure is especially important in the case of our quiescent sample of galaxies, which host minimal gas in the interstellar medium but deep \NaID and \CaII absorption from their evolved stellar populations (dominated by K giant stars).

As in Paper I, we isolate the ISM contribution to the \NaID and \CaII by modeling and removing the stellar continuum using the MaNGA Data Analysis Pipeline (DAP; \citealt{2019AJ....158..231W, 2019AJ....158..160B, 2022ApJS..259...35A}). Briefly, the DAP spatially bins spaxels with Voronoi binning \citep{2003MNRAS.342..345C} to reach $S/N \geq 10$, fits the stellar kinematics with pPXF \citep{2023MNRAS.526.3273C} while masking nebular emission lines, and then refits the spectrum with the stellar kinematics fixed to obtain the best-fit continuum. Throughout this process, we mask the \NaID and \CaII features because of the assumption that these features receive contributions from both the interstellar medium and the stellar photosphere of stars in their host galaxies. We then use the residual, continuum-normalized spectrum to trace ISM gas. Full details of this procedure are given in Paper I.

\begin{figure*}[t]
    \centering
    \includegraphics[width=0.9\textwidth]{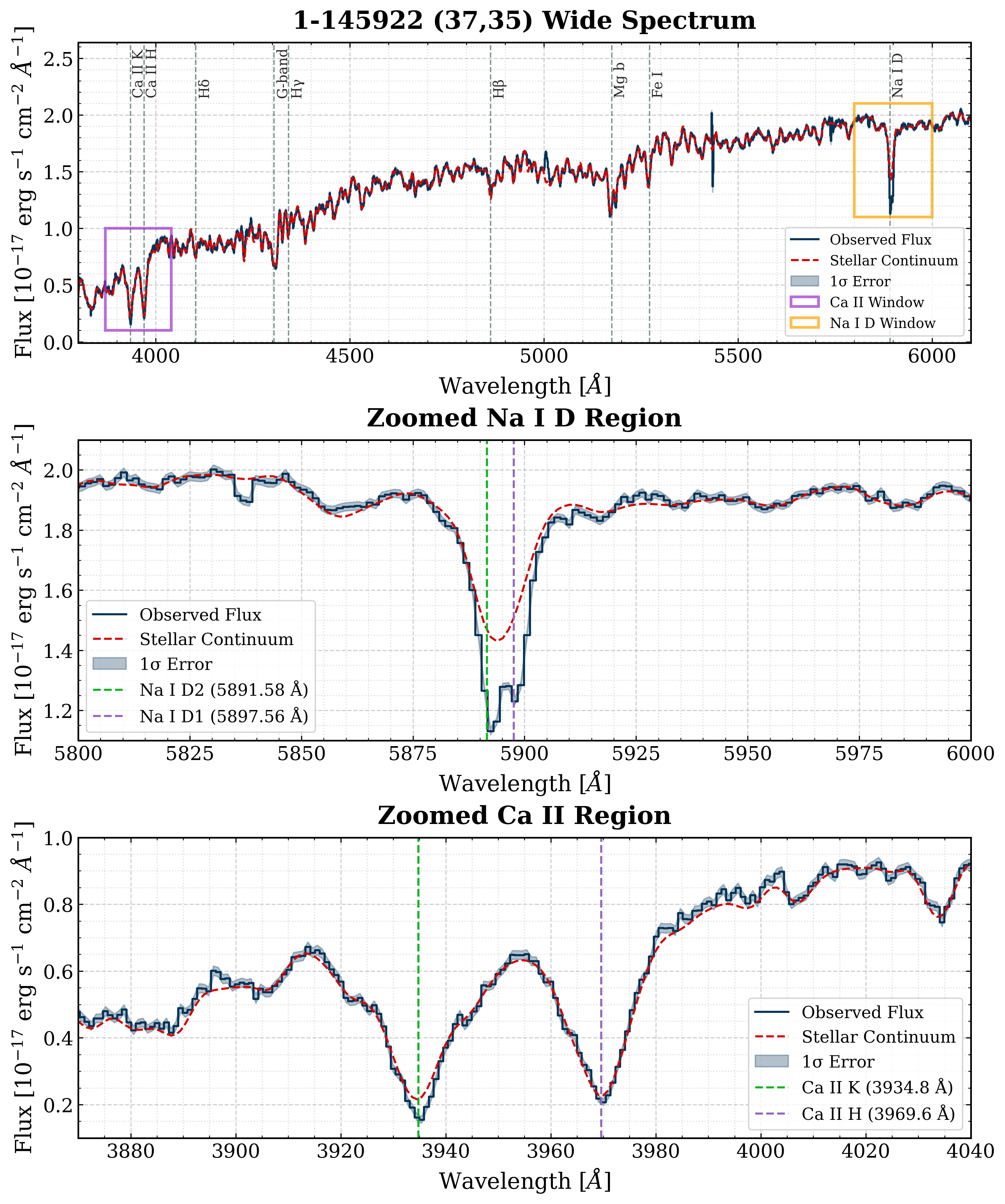} 
    \caption{Example MaNGA spaxel spectrum from the red geyser galaxy 1-145922 for spaxel $(x,y)=(37,35)$. \textbf{Top:} Observed galaxy spectrum over $3800$--$6100\text{ \AA}$ (navy blue) together with the best-fit stellar continuum model from the DAP (red dashed). Prominent spectral features are marked with vertical gray dashed lines. The wavelength ranges shown in the bottom panels (\NaID in the second row and \CaII in the third row) are highlighted by the purple and orange rectangles, respectively. Overall, the stellar continuum model provides a good fit to the observed spectrum. \textbf{Middle:} Zoomed view of the detected \NaID absorption region. The observed spectrum (navy blue) and stellar continuum model (red dashed) are shown, with the locations of the Na I D2 ($\lambda5891.6$; green dashed) and D1 ($\lambda5897.6$; purple dashed) transitions indicated. \textbf{Bottom:} Zoomed view of the \CaII H\&K absorption region, with a detected ISM absorption for the K line. The observed spectrum (navy blue) and stellar continuum model (red dashed) are shown, with the locations of the \CaII K ($\lambda3934.8$; green dashed) and \CaII H ($\lambda3969.6$; purple dashed) transitions marked. In all panels, the shaded region represents the $1\sigma$ uncertainty of the observed flux.}    \label{fig:full_dap_spectrum}
    \end{figure*}

For our stellar templates, we use a hierarchically clustered set of MILES templates (\citealt{2007yCat..83710703S} and \citealt{2019AJ....158..231W}) in which the templates are grouped together based on similar spectral features. The spectrum of one of the galaxies and its best-fit DAP stellar continuum is shown in Figure \ref{fig:full_dap_spectrum}. The DAP does a good job of modeling the observed galaxy spectrum, but to more thoroughly test how well our models fit the observed spectra, we look at absorption features that primarily originate from the stellar photosphere, like the Mg b, g-band, and \CaII triplet features. These absorption features are thought to arise primarily from the stellar photosphere because, unlike \NaID and \CaII H\&K, which are both resonance lines originating from the ground state, they arise from transitions involving excited-state atoms and molecules. Since almost all of the material in the ISM resides in the ground state, the excited-state populations are much more commonly found in stellar photospheres with higher temperatures and higher-density environments.

\begin{figure}[t]
    \centering
    \includegraphics[width=0.46\textwidth]{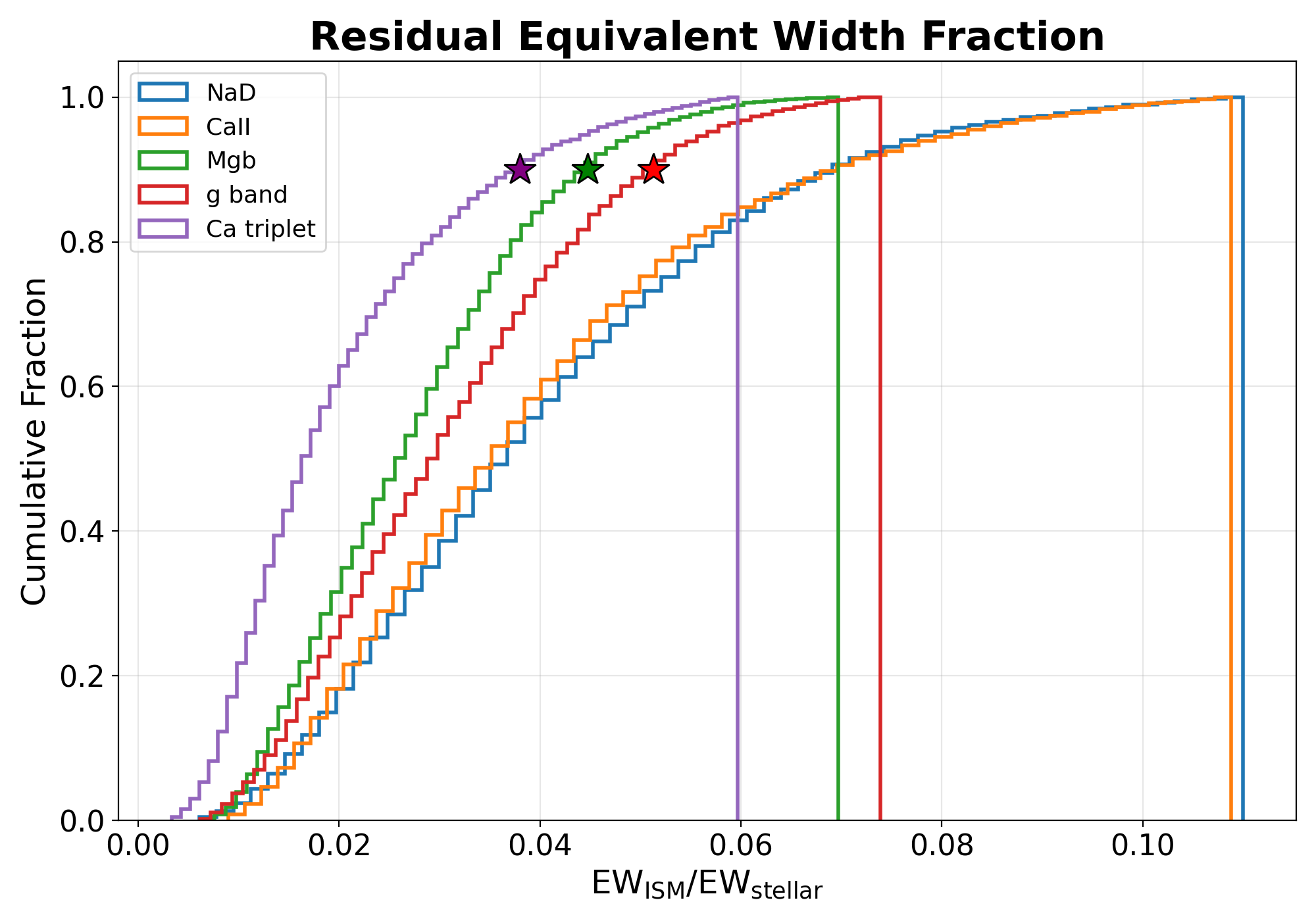} 
    \caption{Cumulative distribution of the residual equivalent width fraction, $EW_{ISM}/EW_{stellar}$, for \NaID (blue), \CaII K (orange), Mg b (green), g band (red), and the \CaII triplet (purple), including both detections and upper limits (since non-detections imply a good stellar template fit). Outliers are removed via the interquartile range (IQR) rule for a better view; the stars indicate the 90th percentile of each stellar absorption feature's distribution. We average the 90th percentile of these three purely stellar features to set a $\sim 5.6\%$ threshold for reliable ISM detection, below which \NaID and \CaII K measurements are excluded from further analysis.}
    \label{fig:poor template fit analysis}
    \end{figure}

Because of this physics, we expect our stellar absorption models to leave no residual for the Mg b, g-band, and \CaII triplet features. We use this as a way to test and remove any poor fits. Using the hierarchically clustered MILES templates, we find that the average ${\text{EW}_{ISM}}/\text{EW}_{*}$ 90th percentile of the data from the stellar absorption features is at $\sim 5.6\%$ (see Figure \ref{fig:poor template fit analysis}). Thus, to be conservative and avoid continuing our analysis on residual features that arise from poor template fits rather than a genuine ISM component, we remove all \NaID and \CaII H\&K absorption with an ISM absorption fraction less than 5.6\%. We also note that many of the spaxels at the outer edges of our galaxy maps (which have been spatially-binned to reach a minimum S/N) exhibit a clear systematic offset between the best-fit model and the observed spectrum. We ensure that we exclude these spaxels in our analysis as well. This leaves us with $\sim30\%$ of the data, where we can confidently rule out poor template fitting as the source of the observed residual absorption.

\begin{figure}[t]
    \centering
    \includegraphics[width=0.48\textwidth]{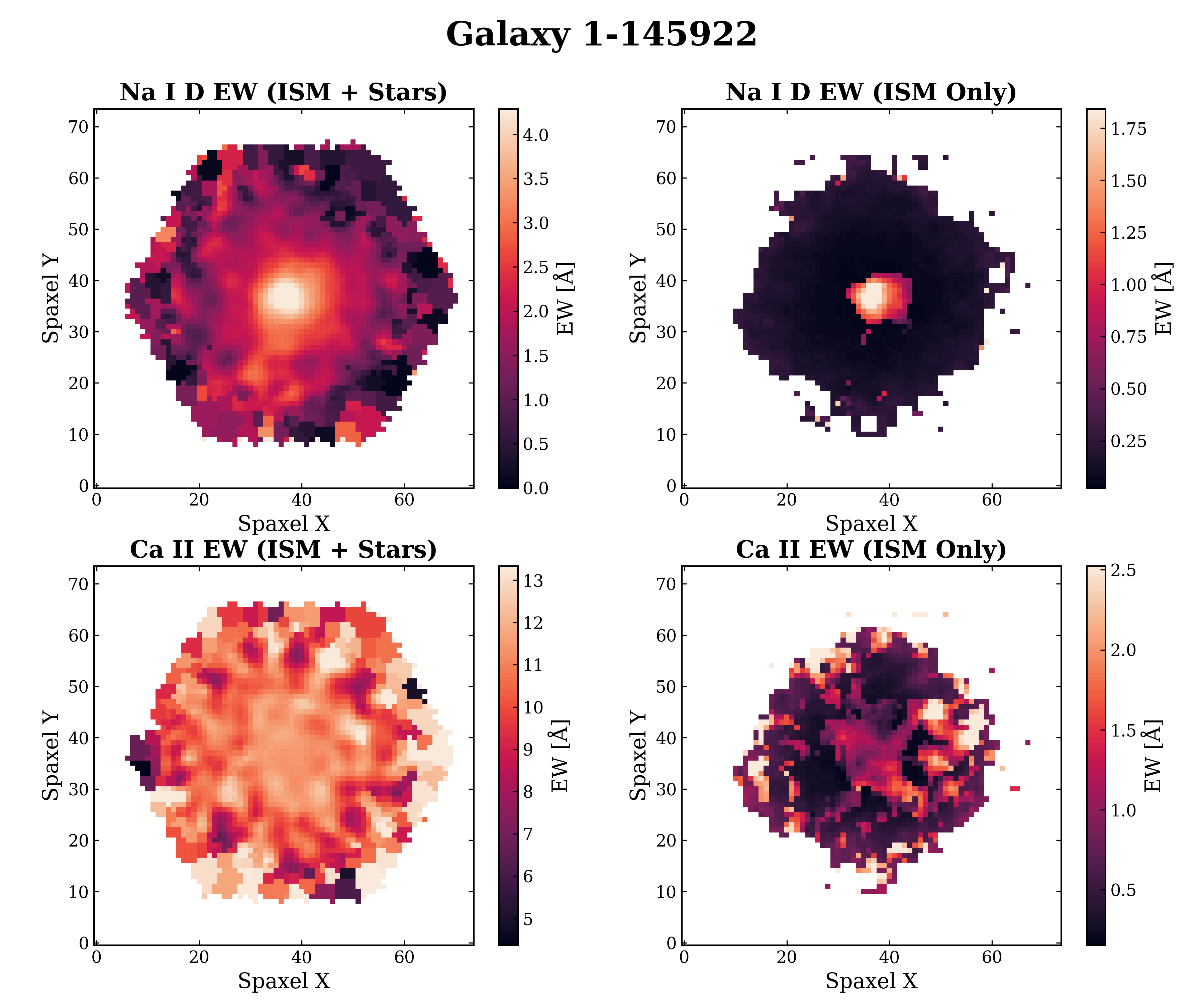} 
    \caption{Equivalent width maps of \NaID and \CaII K absorption before and after removing the stellar absorption contribution from the DAP for an example red geyser galaxy. The left panels show the total observed absorption (interstellar medium + stellar absorption), while the right panels show the same maps after removing the best-fit stellar continuum. The noisy outer spaxels of the maps have been removed because of poor stellar continuum fits.}
    \label{fig:before_after_dap_maps}
\end{figure}

Figure \ref{fig:before_after_dap_maps} shows the measured equivalent width before and after removing stellar contributions from the \NaID and \CaII absorption features. In these maps, it is clear that stellar absorption removal has a significant effect on the resultant maps. 

For the remainder of our analysis, we exclude spaxels with insufficient ISM residuals ($<5.6\%$); then, we take the observed spectrum, as shown in Figure \ref{fig:full_dap_spectrum}, and divide it by the best-fit stellar continuum. These residuals trace cool and warm gas in the ISM, which are the foundations of our analysis.

\subsection{Gaussian Fitting}
\label{subsection: gaussian fitting}
Having removed the stellar contributions to the \NaID and \CaII absorption features, we now seek to model the residual using Gaussian fits and then extract the velocity and velocity dispersion.

To do so, at each spaxel, the normalized flux is modeled using the following expression: 
\begin{equation}
F(\lambda) = 1 - \left[ G_{\mathrm{B}}(\lambda) + G_{\mathrm{R}}(\lambda) \right],
\end{equation}

where each Gaussian component is defined as

\begin{equation}
G(\lambda) = A \, \exp \left[ -\frac{1}{2} 
\left( \frac{\lambda - \left(\lambda_{0} + \Delta\lambda\right)}{\sigma} \right)^{2} \right].
\end{equation}

Here, $A$ is the amplitude, $\sigma$ is the standard deviation (related to the velocity dispersion of the absorbing gas), and $\lambda_{0}$ is the rest-frame wavelength of the transition. For \NNI, $\lambda_{\mathrm{B}} = 5891.58 \,\text{\AA}$, $\lambda_{\mathrm{R}} = 5897.56 \,\text{\AA}$. On the other hand, for \CaII, because of the H$\epsilon$ emission contamination at $3971.2\text{ \AA}$, we only focus on the K line at $\lambda_{\mathrm{B}} = 3934.78 \,\text{\AA}$ and implement a single Gaussian fit on the blue side.

Using these Gaussian fits, we then seek to extract the velocity of these gas clouds. The wavelength shift is defined as \begin{equation}
    \Delta\lambda = \frac{v_{offset}}{c}\times \lambda_0,
\end{equation}
where $c$ is the speed of light, and $v_{offset}$ is the line-of-sight velocity of the gas relative to the mean galaxy redshift, both in $\text{km s}^{-1}$. Since these are ``down-the-barrel" measurements, the underlying stellar continuum acts as the background light, and the foreground cool gas along our line of sight absorbs some of that light, imprinting its own velocity onto that light as absorption. Since we are primarily probing gas in front of the galaxy (with most of the stars in the background), redshifted (positive) gas velocities indicate inflowing motion. In contrast, blueshifted (negative) velocities indicate outflows (i.e., gas moving toward us).

For the double-Gaussian fits of the \NNI, we put a constraint so that the velocity offset and dispersion of the two Gaussians are equal to one another. Additionally, for the \NNI, we constrain the amplitudes of the two Gaussians so that the amplitude of the D1 line at $\lambda_{D1}=5897.56 \text{ \AA}$ is between 0.5 and 0.99 times the amplitude of the D2 line at $\lambda_{D2}=5891.58\text{ \AA}$. The reason for this constraint is that when the gas cloud is in optically thin conditions, the D2 line absorbs roughly twice as much light as D1, given the oscillator strength ratio of $\frac{f_{D2}}{f_{D1}}=\frac{2}{1}$. Therefore, within this regime, the D2 component has a line strength twice that of D1. However, as we move toward the optically thick conditions, the ratio of their strengths approaches 1. 

\begin{figure}[t]
    \centering
    \includegraphics[width=0.48\textwidth]{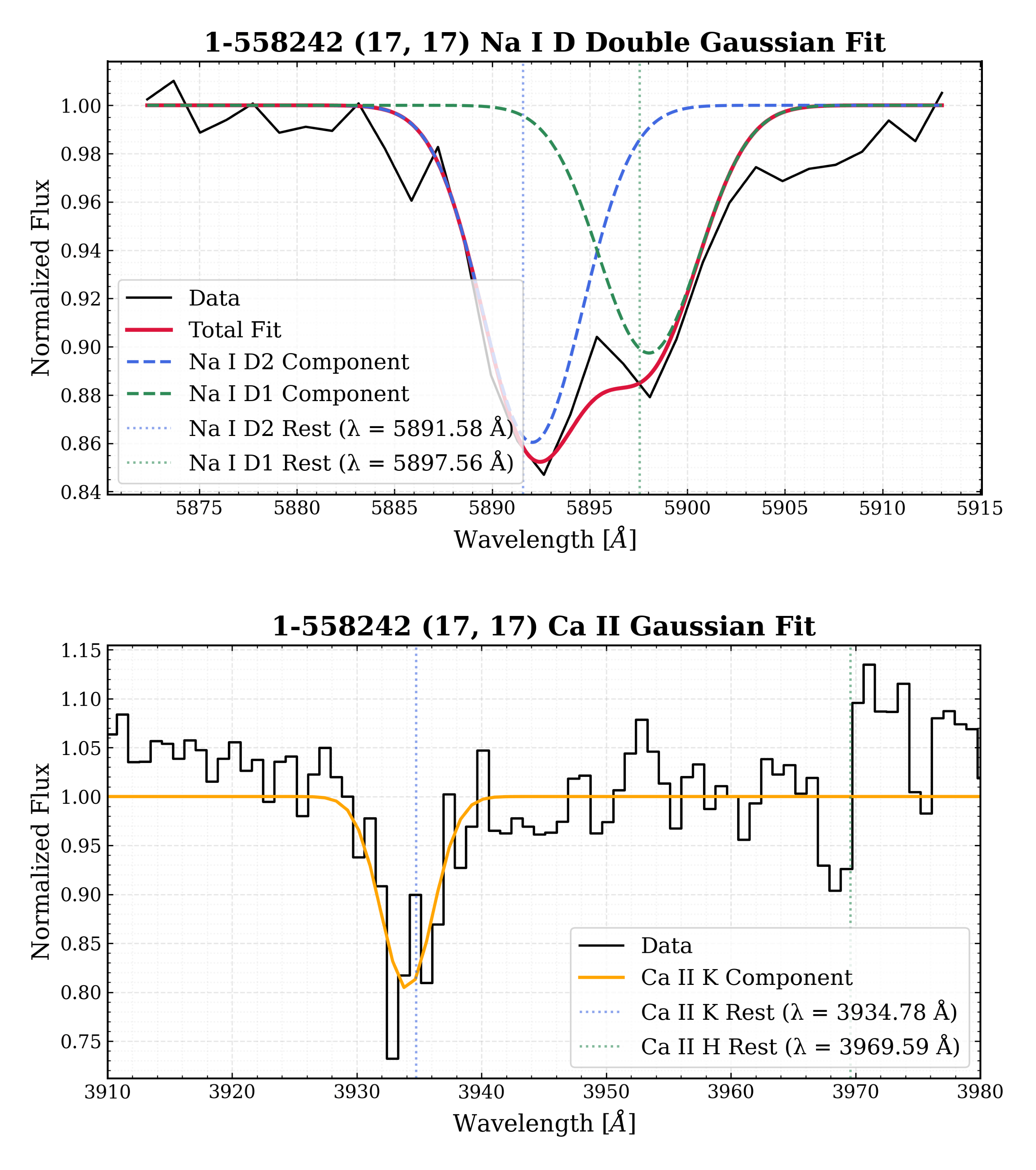} 
    \caption{Example Gaussian fits for the \NaID and \CaII K absorption features in a single MaNGA spaxel. The observed spectrum is shown in black across both panels. In the top row, the \NaID doublet is fit using a double-Gaussian model whose components are shown in green (D1) and blue (D2). The red line represents the composite of the two Gaussian fits as the final \NaID model. The second row shows the \CaII K feature (black) with a single-Gaussian fit model in yellow. We note that the second row does not show the full wavelength range over which the Gaussian fitting is implemented. This is done for better visibility of the \CaII absorption feature. In the Appendix, we describe how we have treated the \CaII H feature, which is filled in by the nearby H$\epsilon$ emission line.}
    \label{fig:gaussian fits}
\end{figure}

Examples of our Gaussian fits for the \NaID and \CaII are shown in Figure \ref{fig:gaussian fits}. The figure demonstrates that the Gaussian models accurately reproduce the residual \NaID and \CaII absorption profiles, allowing us to extract the velocity offset $v_{offset}$ and the dispersion $\sigma$ of the gas clouds from the fits. We note that instead of the velocity dispersion, $\sigma$, we use the full width at half maximum, W50, as our measure of line broadening.

An important step in measuring the line width of the absorption features is to take into account instrumental broadening because the spectrograph has a finite resolution and can smear the signal, so we have to remove that instrumental contribution for a true, intrinsic signal. To do so, we convert our observed $W50_{\text{obs}}$ to $W50_{\text{true}}$ using the following equation:
\begin{equation}
    W50_{\text{true}} = \sqrt{W50_{\text{obs}}^2 - W50_{\text{inst}}^2} ,
\end{equation}
where the $W50_{\text{inst}}$ corresponds to the instrumental broadening in velocity units. MaNGA's spectral resolution is $R\sim2000$ near \NaID and $\sim$1400 near \CCII. Since $R=\frac{\lambda}{\Delta\lambda}$, where $\lambda$ $\approx 5900\text{\AA}$ or $3900\text{\AA}$ and $\Delta\lambda$ is the full width at half maximum of the spectroscopic line-spread function at that $\lambda$, we re-arrange the equation and convert our units, getting $\Delta\lambda_{NaD}\approx165\text{ km s}^{-1}$ and $\Delta\lambda_{CaII}\approx214\text{ km s}^{-1}$. 

Throughout our analysis, we normalize this ISM $W50_{true}$ by the stellar $W50_*$ at the same spaxel to understand how the gas is moving relative to the stars under the same gravitational potential. These $W50_*$ values are measured through the stellar template fits and are also corrected for instrumental broadening.

\subsection{Uncertainties \& Quality Checks of the Fits}
\label{subsection:uncertainties and quality checks}
We evaluate our fits through a Monte Carlo error analysis. We first estimate the noise present in the continuum by calculating the standard deviation, $\sigma$, of the continuum on the blue and red sides of each absorption profile such that they exclude the feature itself. Then, across the full wave range, where we perform Gaussian fitting, we randomly perturb the data points by adding to each point a random value between $-\sigma$ and $+\sigma$. We repeat this process 50 times for each spaxel, each round measuring the velocity offset and velocity dispersion. Lastly, we take the standard deviation of these measurements as the uncertainties for our kinematics measurements. 

We find velocity offset uncertainties of $\sim 14.6 \text{ km s}^{-1}$ for red geysers and $\sim 26.6 \text{ km s}^{-1}$ for the control sample in \NNI, and $\sim29\text{ km s}^{-1}$ and $\sim34.4\text{ km s}^{-1}$, respectively, for \CCII. For the line width measurements, we estimate uncertainties of $\sim7.9\%$ and $10.8\%$ for red geysers and control galaxies, respectively, in \NNI. These uncertainties are once again larger for \CCII, at $16.3\%$ for red geysers and $17.8\%$ for the control sample.

To ensure that our analysis only includes robust detections with reliable kinematics measurements, we apply a series of quality cuts to each spaxel after fitting the interstellar \NaID and \CaII absorption. These quality checks arise from visual inspection of many spaxels and their Gaussian fits. Overall, we exclude any spaxel that does not satisfy the following criteria: 
\begin{enumerate}
    \item The ISM absorption fraction has to be greater than $\sim5.6\%$ of the total observed absorption. This threshold corresponds to the 90th percentile of the measured ``ISM" fraction of purely stellar absorption features, such as Mg b, g-band, and Ca triplet, for which any nonzero residual is not a real interstellar component but simply residual flux from imperfect stellar template fits. This criterion allows us to only keep the features where we can confidently rule out poor template fitting.
    
    \item The fractional uncertainty for the equivalent width has to be less than $50\%$, while the fractional uncertainty on the measured line width is less than $70\%$.

    \item The uncertainty in the velocity offset measurements has to be less than $45 \text{ km s}^{-1}$, and the absolute value of the velocity offset is less than $350 \text{ km s}^{-1}$.

    \item The standard deviation of the residual spectrum over the fitting window (after masking the absorption profiles) is less than 0.6. This helps us exclude spectra with poor continuum subtraction or excessive noise.

    \item The measured absorption line width is less than 3.5 times the stellar line width at the same spaxel's location, removing unphysically broad measurements that likely arise from poor fits.
    
\end{enumerate}

Spaxels that fail the above criteria are removed from our data and excluded from all subsequent analyses. 

Throughout this paper, for \CCII, we only fit the K line ($\lambda$ 3974.78 \AA) because the H line ($\lambda$ 3969.59 \AA) is contaminated by the adjacent H$\epsilon$ emission line. In the Appendix we show that we can remove the H$\epsilon$ emission and recover the presence of \CaII H. However, the resulting uncertainties mean that we cannot reliably extract meaningful measurements. Thus, all the subsequent analysis is based on \CaII K.

\subsection{Dust Reddening} \label{subsec:dust reddening}
A large component of our analysis focuses on dust reddening in our sample of quiescent galaxies and its relationship to the cool and warm gas in the ISM. To quantify the amount of dust extinction, we use the Balmer decrement, which is defined as the flux ratio of the H$\alpha$ and H$\beta$ emission lines. This ratio is expected to be 2.86, but dust preferentially attenuates light at shorter wavelengths (specifically the H$\beta$ in the denominator), which causes the observed ratio to exceed the expected value of 2.86. Observing this effect provides evidence for the existence of dust in the foreground. 

Following \citet{2022ApJ...930..160S}, we use our measured Balmer decrement to estimate a color excess,
\begin{equation}
    E(B-V)=\frac{2.5}{k(\lambda_{H\beta})-k(\lambda_{H\alpha})}\log \frac{f(H\alpha)/f(H\beta)}{2.86},
\end{equation}

where $k(\lambda_{H\beta})=3.66$ and $k(\lambda_{H\alpha})=2.52$ are the values of the \citet{1994ApJ...422..158O} reddening curve assuming that the V-band ratio of total to selective extinction, $R_v=3.1$.  To clean our measurements, we impose a S/N cut of 5 and remove the spaxels with negative $E(B-V)$ values.

With the \NaID and \CaII absorption-line kinematics measured, the dust reddening maps constructed, and unreliable spaxels removed, we now turn to our analysis.

\section{\textbf{Results}} \label{sec:results}

\subsection{Larger Gas Reservoirs Traced by \CaII}
\label{subsection:ISM Gas Reservoir Size}
We begin our analysis by comparing the sizes of the ISM gas reservoirs traced by \NaID and \CaII in red geysers and the control sample. These measurements are based on our detections of interstellar gas traced by \NaID and \CaII in different spaxels. First, in a simple comparison of the number of galaxies with sufficient ($\geq 5$) detections, we find the following results, which are summarized in Table \ref{tab:detection_fractions}:
\begin{enumerate}
    \item For \NNI, red geysers have a detection fraction that is $\sim1.9$ times larger than the control sample (60/140 vs. 31/140).

    \item For \CCII, red geysers have a larger detection fraction (93/140) than the control sample (77/140).

    \item On average, the detection fraction of \CaII is $\sim 1.9$ times larger than that of the \NNI.
\end{enumerate}

Overall, there are more galaxies with detections of \CaII compared to \NNI, and more red geysers host sufficient gas compared to the control galaxies. 

\begin{table}[ht]
    \centering
    \caption{Detection rates of \NaID and \CaII in red geysers and the control sample ($> 5$ spaxels in a galaxy).}
    \label{tab:detection_fractions}
    \begin{tabular}{|l|c|c|}
        \hline
         & Red Geysers & Control Sample \\
        \hline
        \NaID  & 60 ($43\%$) & 31 ($22\%$) \\
        \hline
        \CaII & 93 ($66\%$) & 77 ($55\%$)\\
        \hline
    \end{tabular}
    \tablecomments{On average, the detection fraction of \CaII is $\sim 1.9$ times larger than that of \NNI. Similarly, the detection fraction of red geysers is $\sim 1.4$ times larger than that of the control galaxies.}
\end{table}

\begin{figure}[t]
    \centering
    \includegraphics[width=0.48\textwidth]{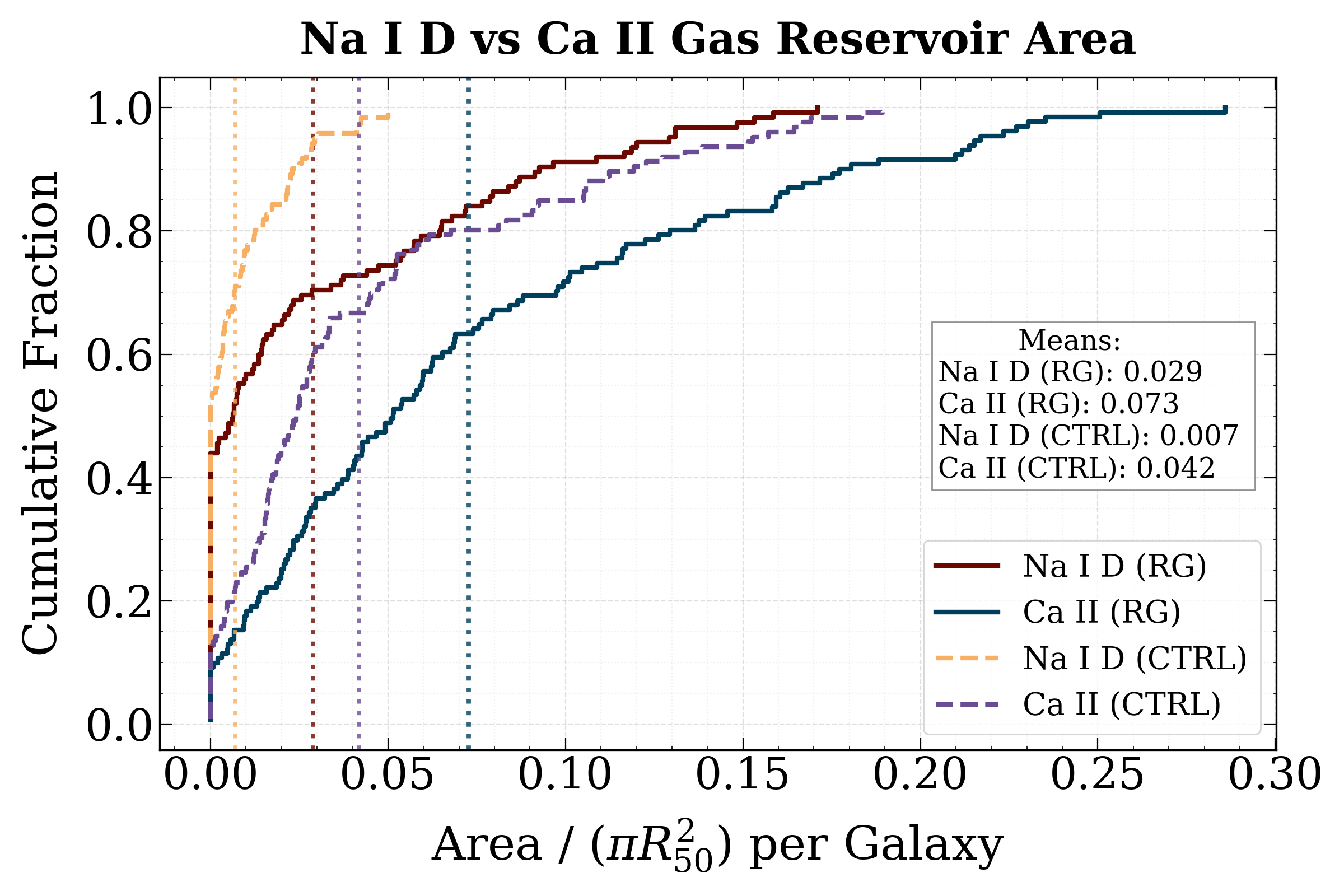} 
    \caption{Empirical cumulative distribution function (ECDF) for the gas reservoir areas across each sample of galaxies, including \NaID in red geysers (solid dark red), \CaII in red geysers (solid dark blue), \NaID in the control galaxies (dashed orange), and \CaII in control galaxies (dashed purple). The total area of each galaxy is normalized by its half-light radius $R_{50}$ for a better comparison across galaxies of different sizes. The vertical dashed lines correspond to the mean values of each distribution. Across both samples, there are larger reservoirs of \CaII detections compared to \NNI, with mean values of $\sim 0.073 \text{ vs. } 0.029$ in red geysers and $\sim 0.042 \text{ vs. } 0.007$ in the control sample. Additionally, compared to the control sample, red geysers systematically host larger ISM gas reservoirs. Outliers are removed using the interquartile range (IQR) rule for better visibility.}
    \label{fig:ISM gas reservoir comparison}
\end{figure}

Following this analysis, we quantify the size of the reservoirs by summing up the area of the detected spaxels for each galaxy. To account for the size of each galaxy in our comparisons, we normalize each measured area by its host galaxy's half-light area, $\pi R_{50}^2$. Figure \ref{fig:ISM gas reservoir comparison} shows that the detected \CaII reservoirs are consistently larger across both samples compared to the \NNI. The average normalized \CaII reservoir covers $7.3\%$ of the galaxy half-light area in red geysers, compared to $\sim2.9\%$ for \NNI. Similarly, in the control sample, the average \CaII reservoir covers $\sim 4.2\%$ of the half-light area, compared to a mean of $0.7\%$ for \NNI. Our Mann-Whitney U test confirms that the difference between the \CaII and \NaID is statistically significant with $p=1.3\times 10^{-7}$ for red geysers and $p=2.1\times 10^{-13}$ for the control sample. Comparing the two galaxy populations shows that red geysers systematically host larger ISM gas reservoirs than the control sample. Specifically, the average \CaII reservoir is $\sim 1.7$ times larger in red geysers, while the mean \NaID reservoir in red geysers is $\sim 4.1$ times larger than in the control sample.

\begin{figure}[t]
    \centering
    \includegraphics[width=0.48\textwidth]{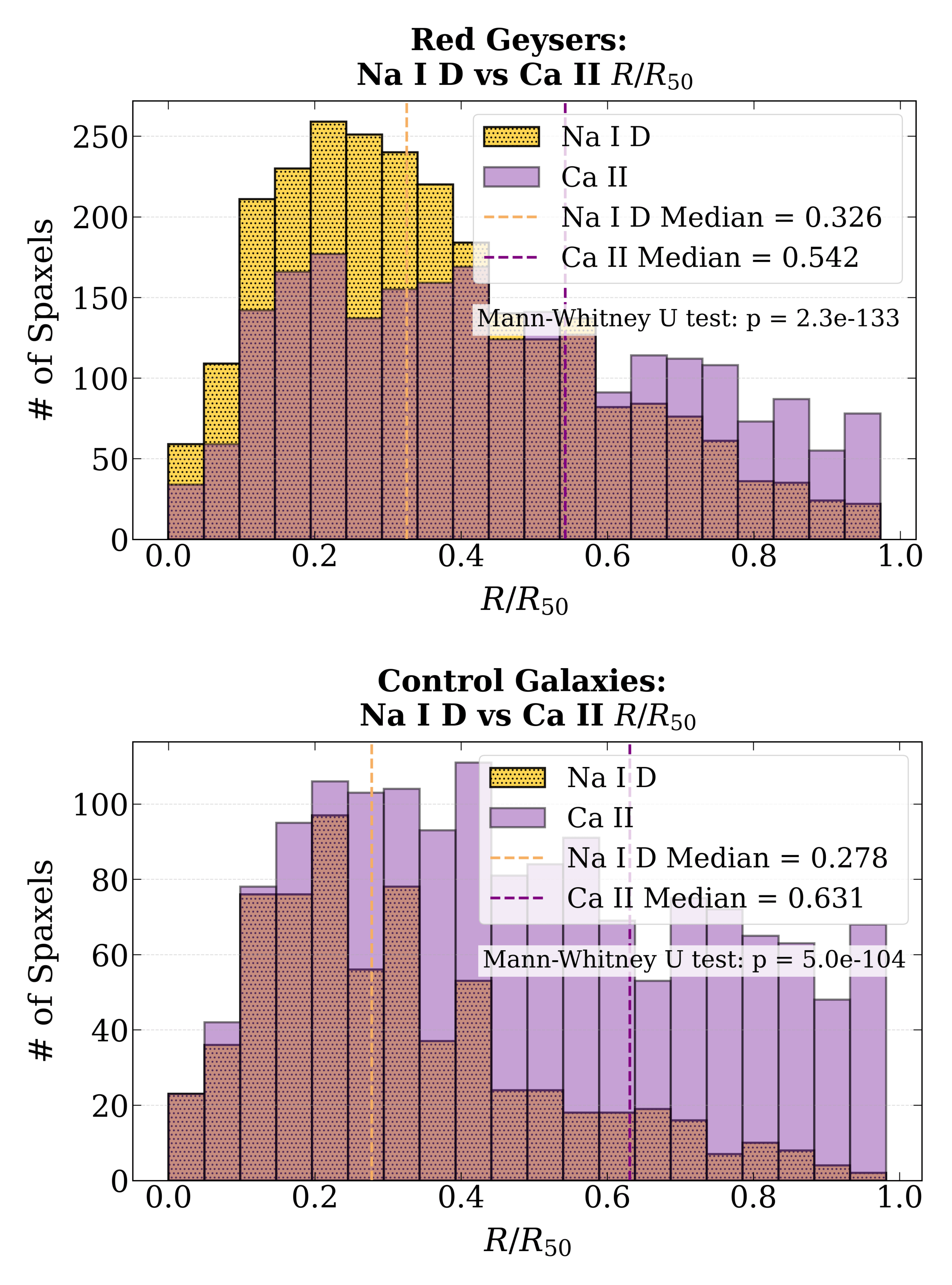} 
    \caption{Radial distributions of cool, neutral (\NNI, yellow) and warm ionized (\CCII, purple) gas in red geysers (top) and the control sample (bottom). Vertical dashed lines indicate the median of each distribution. In both samples, \CaII detections extend significantly farther than \NaID detections, with median values of $R_{CaII}/R_{50}=0.542$ vs. $R_{{NaD}}/R_{50}=0.326$ for red geysers and $R_{{CaII}}/R_{50}=0.631$ vs. $R_{{NaD}}/R_{50}=0.278$ for the control sample. Mann--Whitney $U$ tests confirm that these differences are highly significant for both red geysers ($p=2.3\times10^{-133}$) and the control sample ($p=5\times10^{-104}$). In contrast, differences between red geysers and the control sample are comparatively modest, and unlike the gas reservoir sizes, the radial extents do not show a consistent enhancement in red geysers. Outliers have been excluded using the interquartile range (IQR) rule for clarity.}
    \label{fig:Radial distributions}
\end{figure}

To further understand the radial extent of these gas detections, we also calculate the distance from the center of each galaxy to the location of the \NaID and \CaII detections, normalized by the half-light radius of the host galaxy. As shown in Figure \ref{fig:Radial distributions}, the \CaII detections extend farther out relative to the \NaID across both the red geyser and control samples. Taking median values we get $R_{CaII}/R_{50}\sim 0.542 \text{ vs. } R_{NaD}/R_{50}\sim 0.326$ for the red geysers and $R_{CaII}/R_{50}\sim 0.631 \text{ vs. } R_{NaD}/R_{50}\sim 0.278$ for the control sample. Indeed, our Mann-Whitney U test confirms that the \NaID and \CaII distributions are different from one another on a statistically significant level, with a $p$ value of $\sim 2.3\times 10^{-133}$ for red geysers and $\sim 5\times10^{-104}$ for the control sample. 

Despite these differences, comparisons between the red geysers and the control sample show a weaker difference between the two samples. Although the \NaID and \CaII radial distributions differ at a statistically significant level (Mann--Whitney $U$ tests; $p_{{NaD}}=1.2\times10^{-2}$ and $p_{{CaII}}=2.2\times10^{-7}$), the change in sizes is modest, and unlike the gas reservoir sizes, the radial distributions do not exhibit a large enhancement in red geysers.

Overall, we detect larger reservoirs of \CaII than \NNI, with red geysers hosting systematically larger ISM reservoirs compared to the control sample of galaxies.

\subsection{Broader Line Widths Traced by \CaII}
\label{subsection:line width comparison}
\begin{figure}[t]
    \centering
    \includegraphics[width=0.48\textwidth]{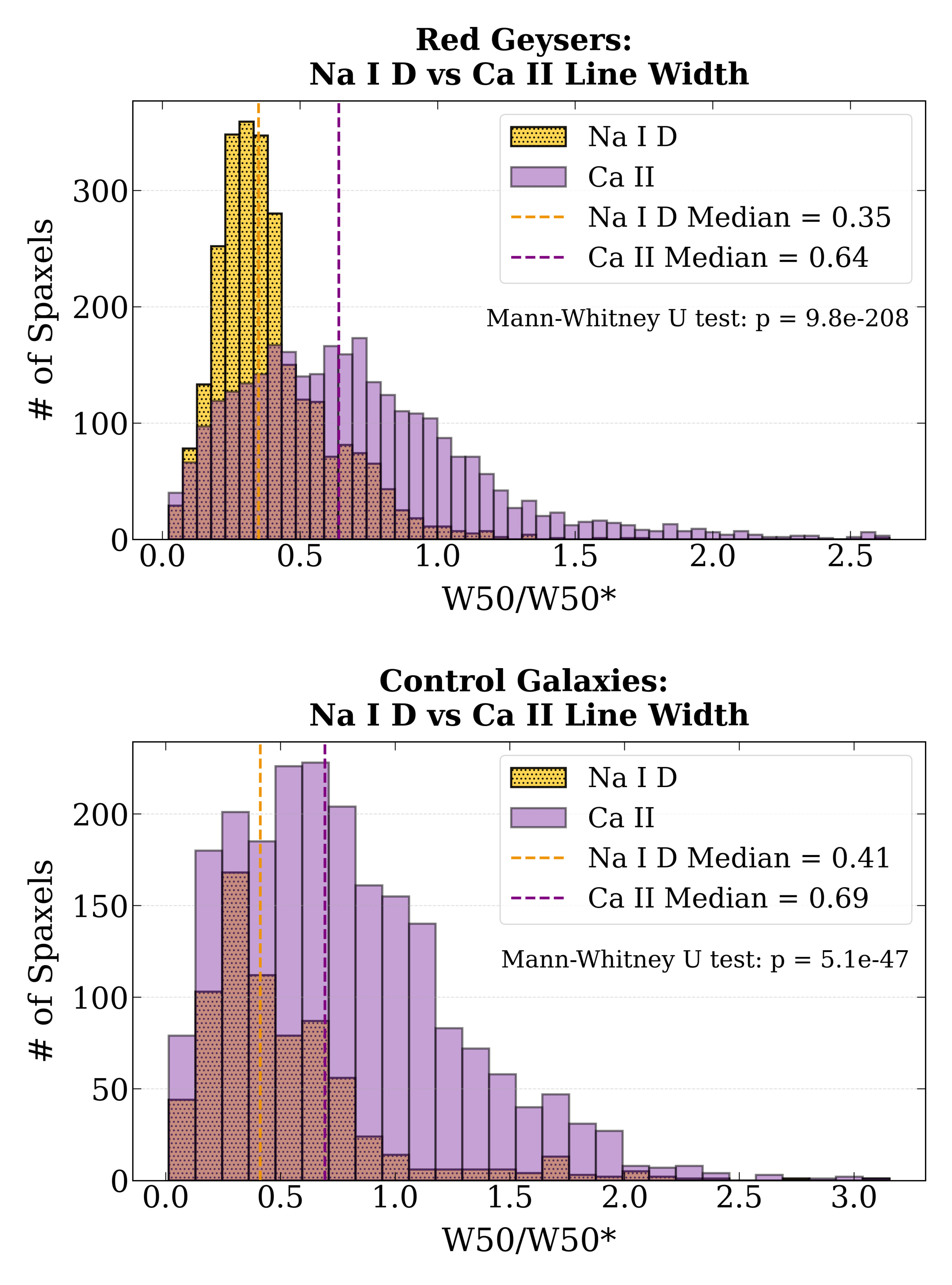} 
    \caption{Distributions of line width measurements $W50/W50_*$ relative to stars for \NaID (yellow) and \CaII (purple) in red geysers (top) and the control sample (bottom). Vertical dashed lines are used to indicate the median values of each distribution. Across both red geysers and the control sample, the \NaID and \CaII lines are narrower than the stellar lines. Also, the \CaII lines are broader than the \NaID lines (median $W_{50}/W_{50}^{*} = 0.64$ vs. $0.35$ in red geysers, and $0.69$ vs. $0.41$ in the control sample; Mann--Whitney U test, $p = 9.8\times10^{-208}$ and $p = 5.1\times10^{-47}$, respectively). When comparing red geysers to the control sample, the control sample shows slightly broader lines than red geysers that are $\sim 1.17$ and $1.08$ times larger for \NaID and \CaII, respectively.}
    \label{fig:Linewidth distributions}
\end{figure}

We next compare the line width measurements of the \CaII to those of the \NaID in both the red geyser and control samples. To gain a better intuition of our dispersion measurements, we normalize our interstellar gas $W50_{ISM}$ values by the stellar $W50_*$ at the same spatial location in the galaxy (same spaxel). Figure \ref{fig:Linewidth distributions} shows two things. First, both the \CaII and \NaID lines are systematically narrower than the stellar absorption lines. This means that the gas must have a different dynamical and physical structure than the stars. 

Second, the \CaII absorption line widths tend to be much broader than the line widths measured for the \NNI. More specifically, the median line width is $\sim 1.8\times $ larger for \CaII than \NaID in red geysers, and $\sim 1.7\times$ larger in the control sample. A comparison of the red geysers to the control sample shows that red geysers host slightly narrower lines overall, and this difference is statistically significant ($p_{NaD}\sim 3\times 10^{-12} $,  $p_{CaII}\sim5\times 10^{-8}$). However, this red-geysers-vs-control difference is smaller than the \NNI-vs-\CaII difference described above: the control sample's \NaID lines are only $\sim 1.17 \times$ broader than in red geysers, while its \CaII lines are just $\sim 1.08\times$ broader.

Overall, \CaII traces gas with larger velocity dispersion than \NNI. Additionally, compared to red geysers, the control sample hosts slightly broader \CaII and \NaID features. Both the \NaID and \CaII lines are narrower than the stellar absorption lines. The stars and the two gas phases must all have different structures from one another.

\subsection{Comparison of Velocities}
\label{subsection:velocity comparison}
\begin{figure}[t]
    \centering
    \includegraphics[width=0.48\textwidth]{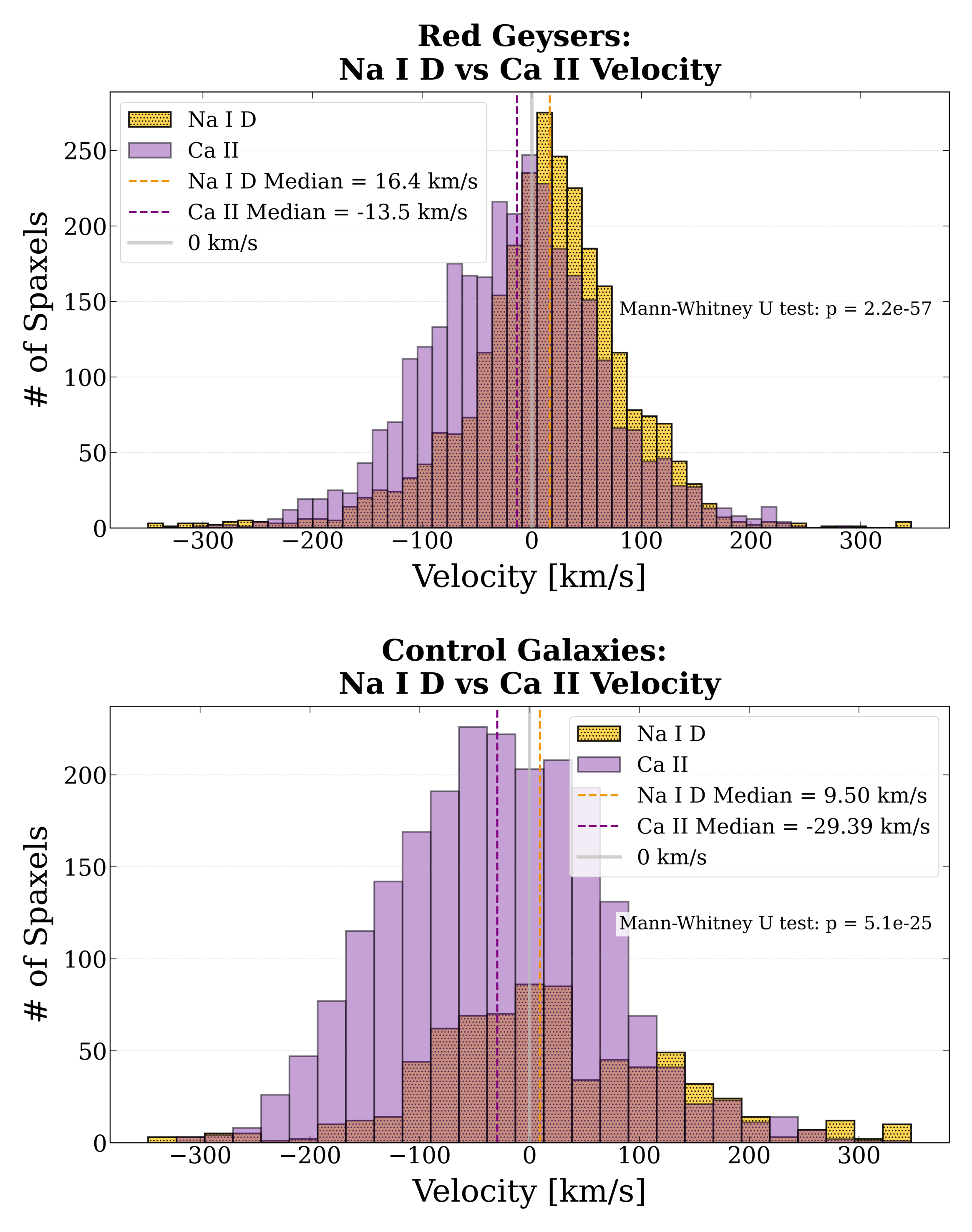} 
    \caption{Distributions of velocity offset measurements $v_{offset}$ relative to the host galaxy for \NaID (yellow) and \CaII (purple) in red geysers (top) and the control sample (bottom). Vertical dashed lines are used to indicate the median values of each distribution. Across both red geysers and the control sample, \CaII seems to be associated with more outflowing gas (median $\sim -13.5 \text{ km s}^{-1}$ in red geysers and $\sim -29.39 \text{ km s}^{-1}$ in control sample), while \NaID is associated with more inflows (median $\sim +16.4 \text{ km s}^{-1}$ in red geysers and $\sim +9.5 \text{ km s}^{-1}$ in control sample). The Mann--Whitney U test yields $p = 2.2\times10^{-57}$ for red geysers and $p = 5.1\times10^{-25}$ in the control sample.}
    \label{fig:velocity distributions}
\end{figure}

Following up on our velocity dispersion comparisons, we next turn to our velocity offset, $v_{offset}$, measurements. We find that in red geysers, the \NaID distribution is mildly redshifted with a median velocity of $+16.4 \text{ km s}^{-1}$, while the \CaII distribution is mildly blueshifted with a median of $\sim -13.5 \text{ km s}^{-1}$ (Figure \ref{fig:velocity distributions}). A similar pattern is observed in the control sample, where \NaID is redshifted at $+9.5 \text{ km s}^{-1}$, and \CaII is blueshifted at $-29.39 \text{ km s}^{-1}$. Across both samples, a Mann-Whitney U test confirms that the differences between \NaID and \CaII velocity distributions are statistically significant ($p=2.5\times 10^{-57}$ in red geysers and $p=5.1\times 10^{-25}$ in the control sample). Within the uncertainties of our measurements, described in Section \ref{subsection:uncertainties and quality checks}, the median values of our velocities seem to suggest that the two absorption features could trace kinematically distinct gas (consistent with the differences we see in the line widths). We note, however, that these velocity distributions are broad, and their average standard deviation is $\sim 90 \text{ km s}^{-1}$, causing large overlaps between the \NaID and \CaII distributions. Therefore, the minor differences in the medians of the distributions of measured velocities should be taken in this context.

\begin{figure*}[t]
    \centering
    \includegraphics[width=0.98\textwidth]{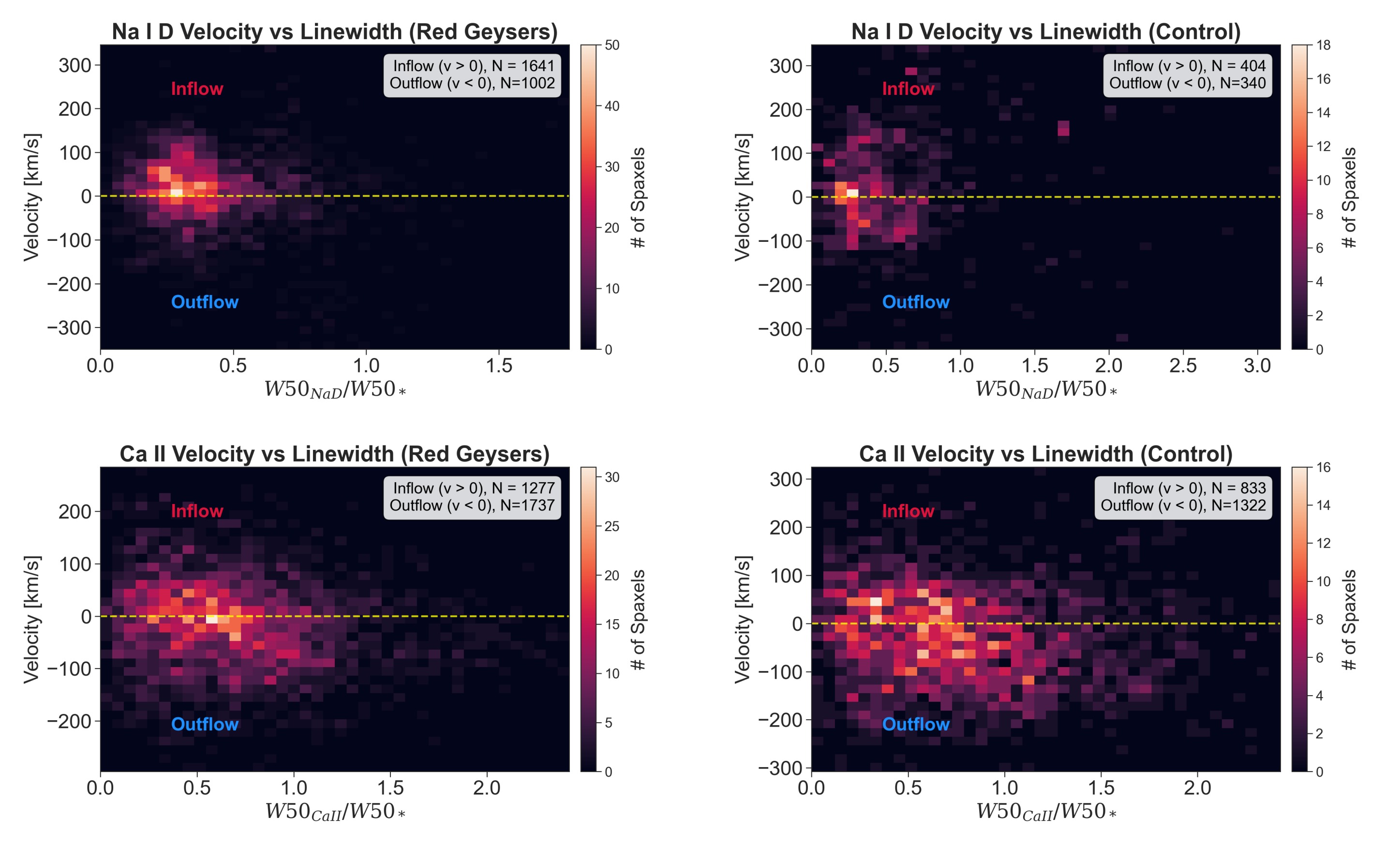} 
    \caption{\NaID and \CaII velocity vs. the normalized line width (relative to the stellar line widths) in red geysers and the control sample. The plots make it clear that the line width measurements of \NaID are limited to $W50_{NaD}/W50_*\lessapprox 1$, but \CaII shows no such limit until $W50_{CaII}/W50_*\sim 2$. Additionally, the largest number of detections occurs in 1) red geysers compared to the control and 2) \CaII compared to \NNI. Lastly, \NaID absorption appears to have a larger fraction of inflowing cool gas compared to \CaII ($\sim 62\%$ vs. $\sim 42\%$ in red geysers, and $\sim 54\%$ vs. $\sim 38\%$ in the control sample). }
    \label{fig:velocity vs linewidth scatter plots}
\end{figure*}

Figure \ref{fig:velocity vs linewidth scatter plots} shows a summary of the kinematics of the gases traced by \NaID and \CaII in red geysers and the control sample. Three trends stand out:
\begin{enumerate}
\item First, \CaII shows systematically larger line widths than \NNI: almost no \NaID detections exceed $W50_{NaD}/W50_{*} >1$ ($\sim 2\%$ in red geysers and $\sim 9\%$ in the control sample), but \CaII shows no such constraint, with its detections continuing well beyond this limit ($\sim 28\%$ and $\sim 20\%$, respectively). This suggests that \NaID primarily traces kinematically colder gas compared to \CCII.

\item Second, \NaID tends to be preferentially associated with positive-velocity inflows, while \CaII shows a better connection with negative velocity outflows. Specifically, in red geysers, $\sim 62\%$ of our \NaID detections are inflowing, compared to only $\sim 42\%$ for \CaII ($\sim 58\%$ outflowing); a similar pattern is seen in the control sample, with $\sim 54\%$ of \NaID detections inflowing vs. $\sim 38\%$ for \CaII ($\sim 62\%$ outflowing). 

\item Another result evident in these plots is that \CaII yields more detections than \NNI, and red geysers host more detections than the control sample. \CaII in red geysers has the largest number of gas detections (3014); this is $\sim 4$ times larger than the \NaID detections in the control sample (744), the lowest of the four. This is consistent with the picture that \CaII traces a larger and more extended gas reservoir than \NNI, and with red geysers hosting more gas overall than the sample of control galaxies.

 \end{enumerate}

\begin{figure}[t]
    \centering
    \includegraphics[width=0.49\textwidth]{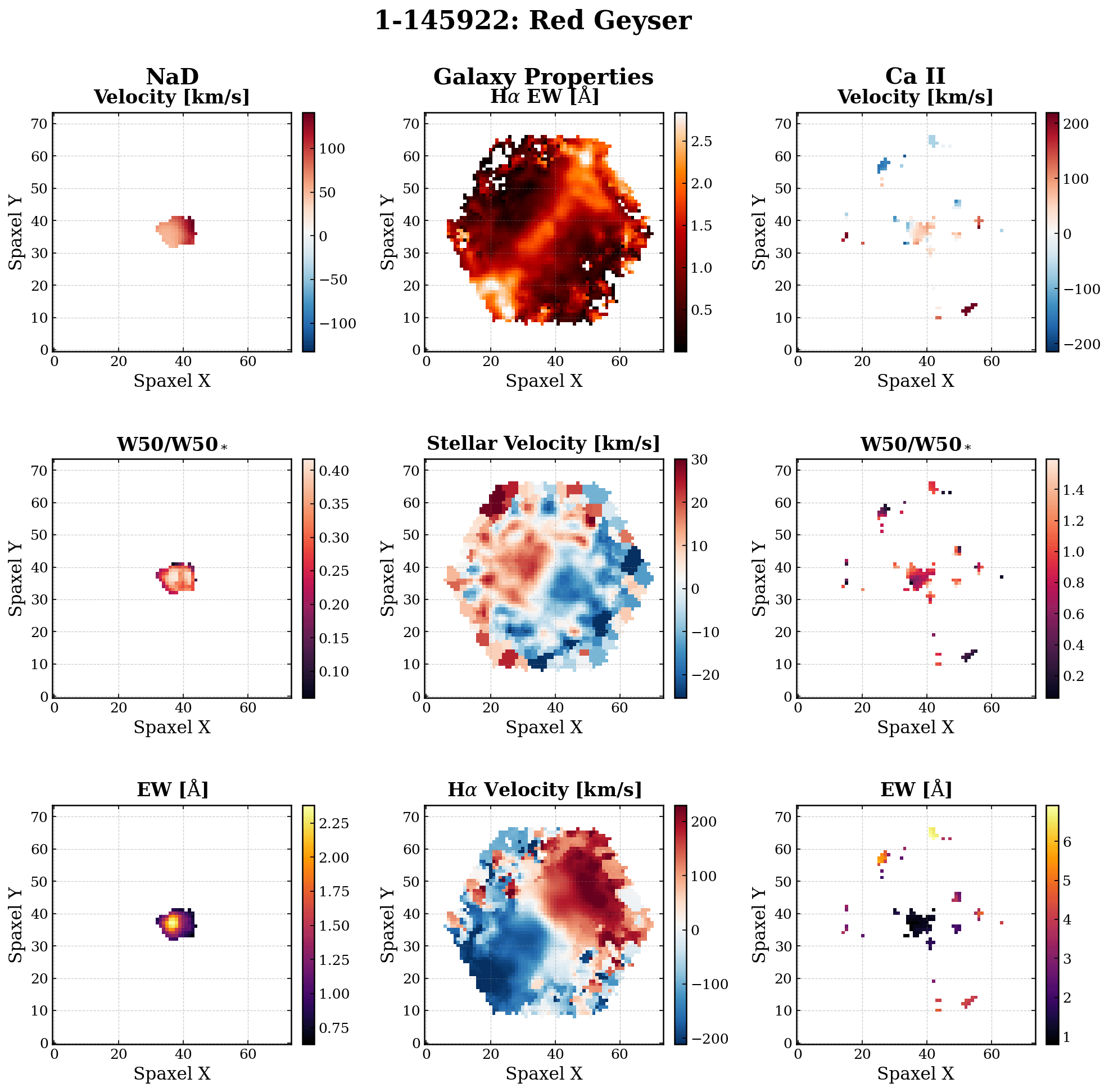} 
    \caption{Spatially resolved maps for an example red geyser galaxy (1-145922). The left and right columns show properties of \NaID and \CCII, respectively, while the central column shows the same galaxy's stellar and H$\alpha$ properties. The bisymmetric feature that characterizes a red geyser is clearly shown in the H$\alpha$ EW map in the middle column.}
    \label{fig:red geyser summary map}
\end{figure}

\begin{figure}[t]
    \centering
    \includegraphics[width=0.49\textwidth]{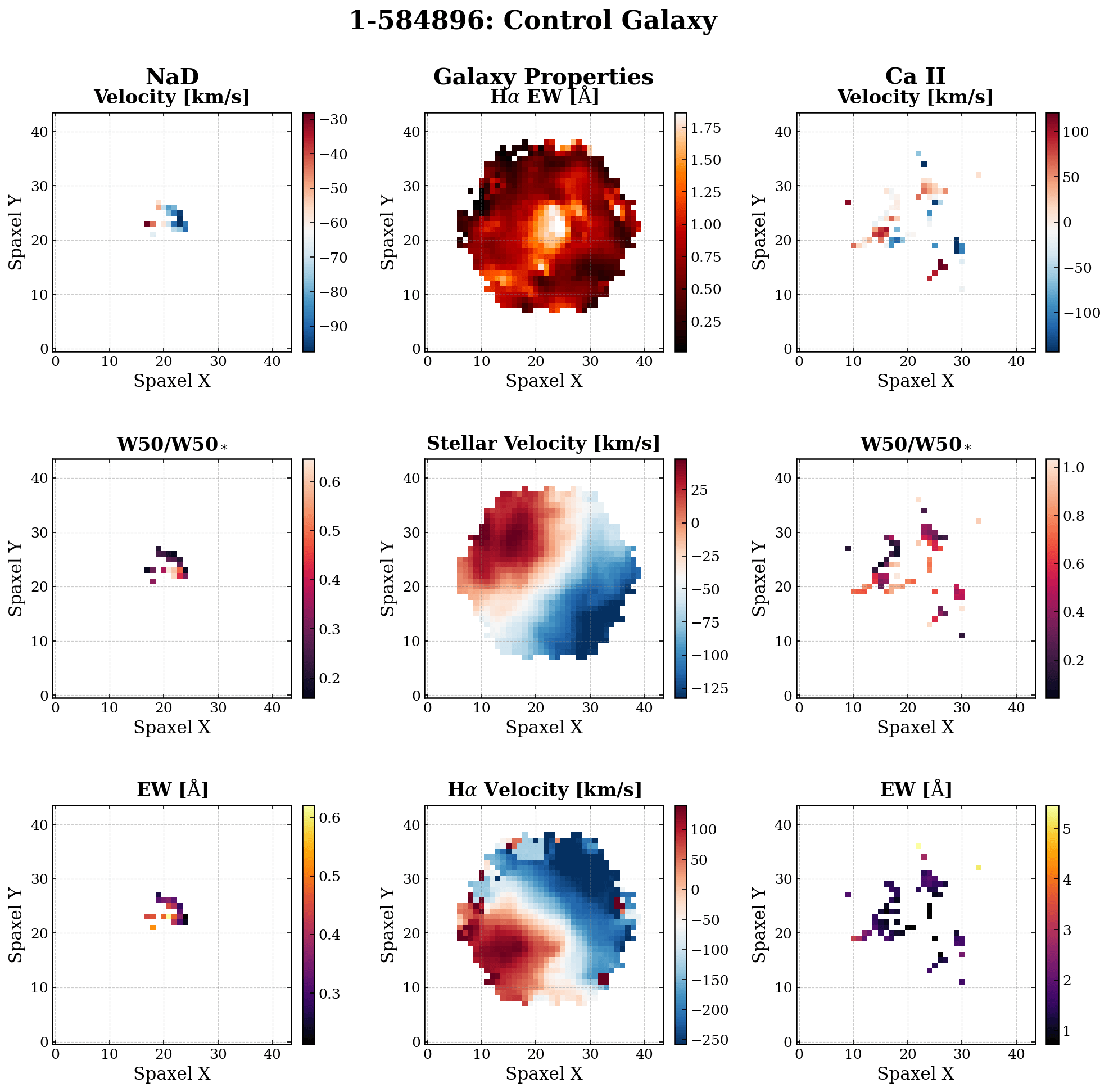} 
    \caption{Spatially resolved maps for an example Control galaxy (1-584896). The left and right columns show properties of \NaID and \CCII, respectively, while the central column shows the same galaxy's stellar and H$\alpha$ properties. Unlike red geysers, the H$\alpha$ EW map does not exhibit a bisymmetric feature.}
    \label{fig:Control summary map}
\end{figure}

Example maps of \NNI, \CCII, and their host galaxy properties, such as H$\alpha$ EW and stellar kinematics for a red geyser and a control galaxy, are shown in Figures \ref{fig:red geyser summary map} and \ref{fig:Control summary map}, respectively. These maps show the radially extended \CaII detections compared to the \NNI, as well as the higher W50/W50* for \CCII.

Overall, in our sample, the majority of the \CaII is outflowing gas, while the majority of the \NaID is inflowing gas.

\subsection{Red Geysers vs. Control: Dust Reservoir Size}
\label{subsection:Red Geysers Host Larger Dust Reservoirs}
Following our analysis of the prevalence and kinematics of cool and warm gas in the interstellar medium of red geysers and the control sample, we now turn to their connection to dust. In this section, we first quantify the size of the dust reservoirs in our sample, comparing red geysers to the control sample, before examining their connection to Na I D and Ca II in Section \ref{subsection:Dependence on Dust for Survival}.

\begin{figure}[t]
    \centering
    \includegraphics[width=0.48\textwidth]{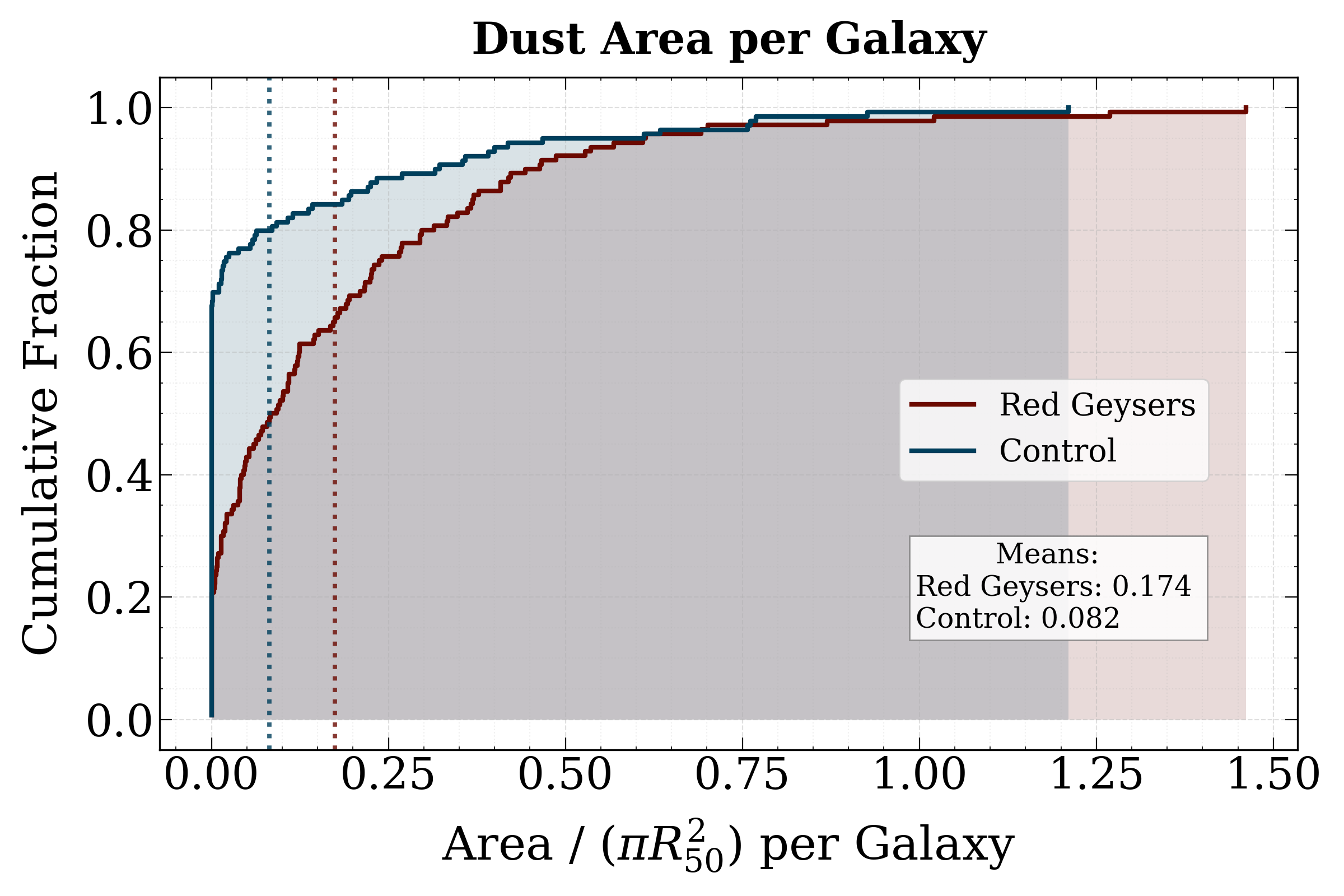} 
    \caption{Empirical cumulative distribution function (ECDF) for dust area per galaxy, normalized by $\pi R_{50}^2$, for red geyser galaxies (dark red) and the control sample (blue). Vertical dashed lines indicate the mean of each distribution (red geysers: 0.174; control: 0.082). Red geysers host larger dust reservoirs than control galaxies.}
    
    \label{fig:dust histogram comparison}
\end{figure}

To quantify the size of the dust reservoirs in each galaxy, we once again sum up the total area covered by our dust detections. We then normalize this area by the effective area $\pi R_{50}^2$ of each galaxy to account for differences in galaxy size, and compare the resulting distributions between the red geyser galaxies and the control sample. Figure \ref{fig:dust histogram comparison} shows that red geysers host dust reservoirs that, on average, are $\sim 2$ times larger than the dust reservoirs of the control galaxies. Additionally, the 16th percentile of dust area is zero in both samples. At the 84th percentile, however, the samples separate: control galaxies have a normalized dust reservoir covering $\sim 15\%$ of their host galaxy's effective area, while in red geysers this fraction more than doubles to $\sim 37\%$. Our Mann-Whitney U test yields a $p$ value of $\sim 2\times 10^{-12}$, confirming that the difference between the two distributions is statistically significant. 

\begin{figure}[t]
    \centering
    \includegraphics[width=0.48\textwidth]{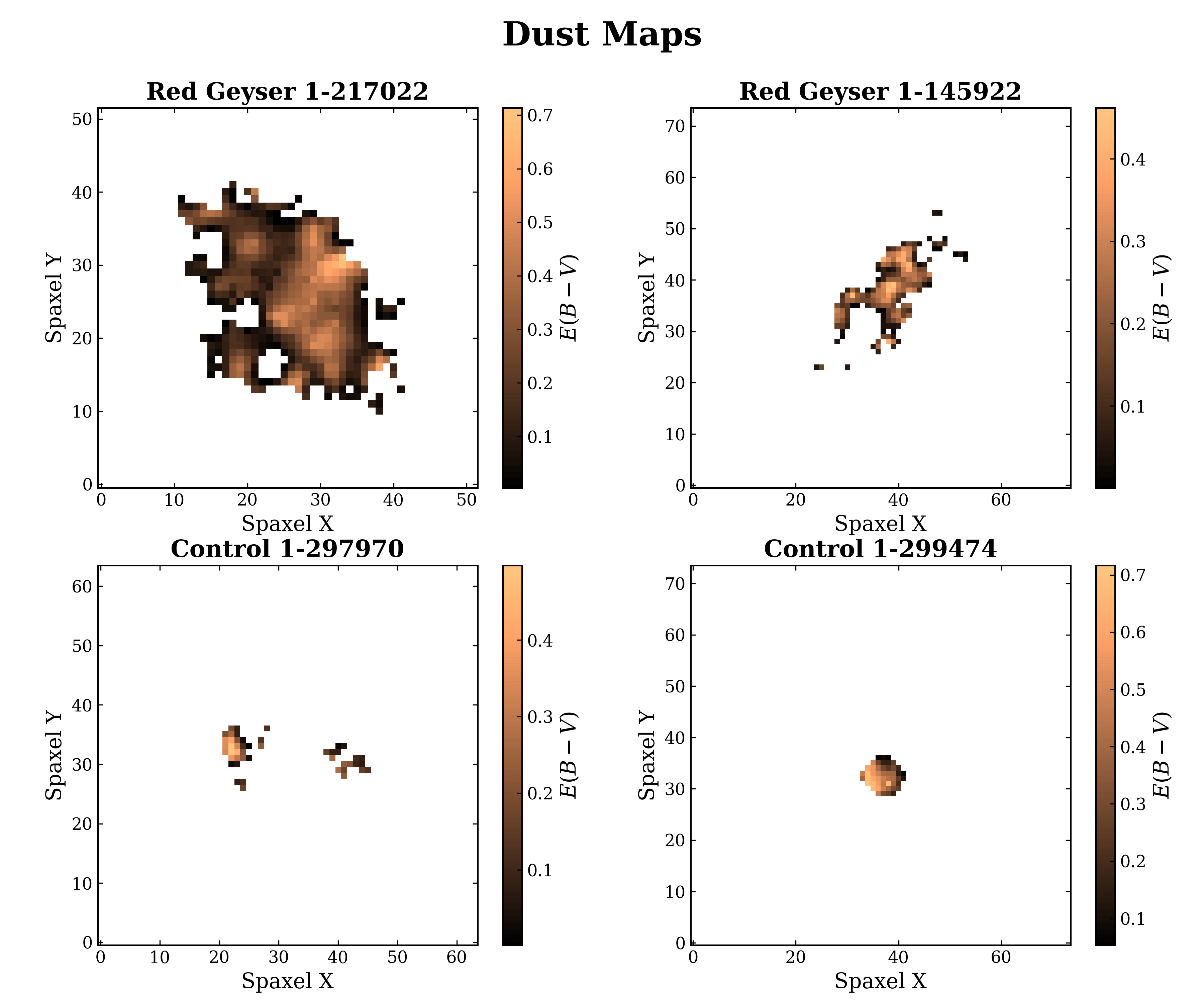} 
    \caption{Example dust maps of two red geyser galaxies (top row) and control galaxies (bottom row), showing the distribution of $E(B-V)$ across each galaxy's spaxels. Red geysers show substantially larger dust reservoirs compared to the control sample.}
    \label{fig:dust maps}
\end{figure}

Example maps of the dust reservoirs in two red geysers and two control galaxies are shown in Figure \ref{fig:dust maps}. Red geysers host larger and more extended dust reservoirs compared to the control sample.

\subsection{Dust Shielding of ISM Gas}
\label{subsection:Dependence on Dust for Survival}

\begin{figure}[t]
    \centering
    \includegraphics[width=0.49\textwidth]{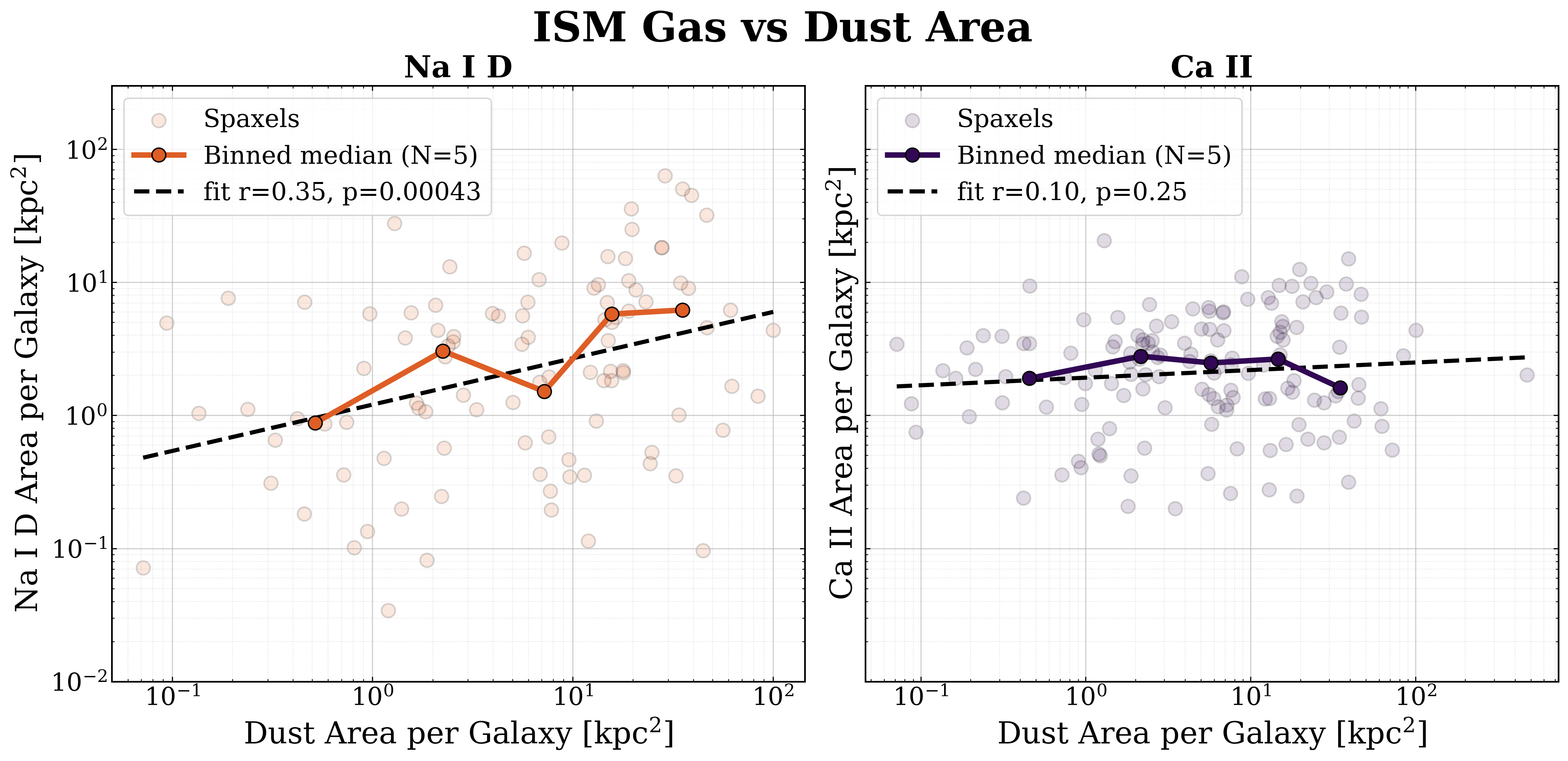} 
    \caption{\NaID (orange) and \CaII (purple) reservoir area vs. dust area for red geysers and the control sample (combined). Individual galaxies are shown with lighter-colored points, while the data are divided into five bins and the median $x$ and $y$ values of each bin are shown with darker symbols to highlight the underlying trends. Black dashed lines indicate the best-fit linear relations. The Pearson correlation coefficient, $r$, and corresponding $p$ value are also included in each panel.}
    \label{fig:NaD and CaII vs dust correlation}
\end{figure}

Now that we have mapped the dust reservoirs of red geysers and the control sample, we seek to understand their connection to the gases traced by \NaID and \CaII absorption. To examine this connection, we start by analyzing the relationship between the size of the dust reservoirs and gases traced by \NaID and \CCII. Figure \ref{fig:NaD and CaII vs dust correlation} shows that there is a modest ($r=0.35$) and statistically significant ($p=4.3\times10^{-4}$) correlation between the size of the \NaID and dust reservoirs. However, this correlation vanishes when we compare \CaII to the dust reservoirs ($r=0.1$ and $p=0.25$).

Furthermore, we compare the radial extent of the dust reservoirs with that of the interstellar gas. For each galaxy, we define the maximum radial extent of the gas using the median distance of the five farthest detections (spaxels) from the center to reduce the influence of outliers. We then compare these gas extents with the corresponding radial extent of the dust to see whether the dust reservoirs are associated with the gas reservoirs in the ISM. In our sample, only the following fraction of galaxies host sufficient data for analysis:
\begin{enumerate}
\item \textbf{\NNI:} $\sim 33\%$ (91/280)
\item \textbf{\CCII:} $\sim 61\%$ (170/280) 
\item \textbf{Dust:} $\sim 46\%$ (130/280)
\end{enumerate}

\begin{figure*}[t]
    \centering
    \includegraphics[width=0.98\textwidth]{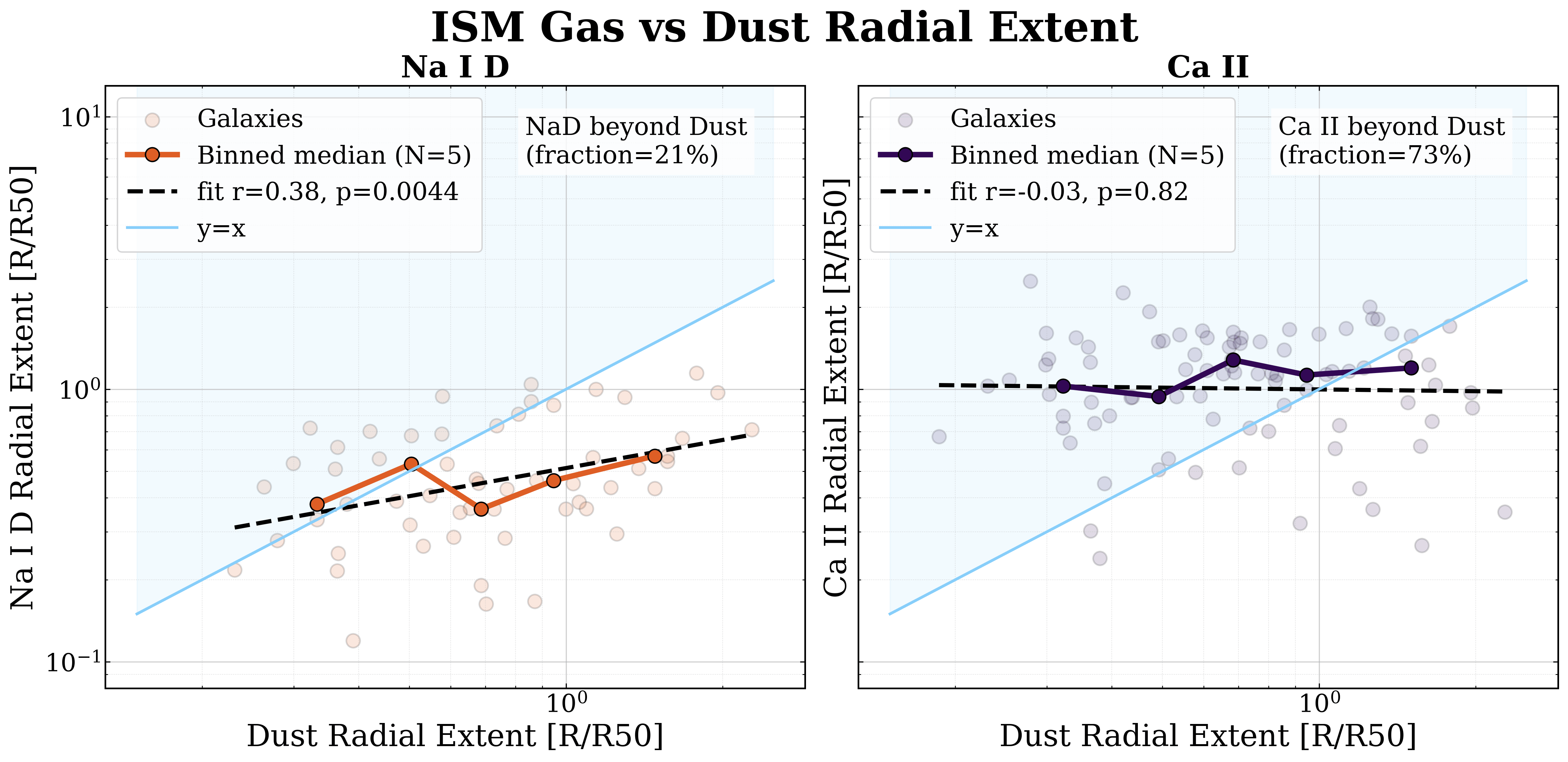} 
    \caption{Radial extent of \NaID (orange) and \CaII (purple) detections vs. dust for red geysers and the control sample combined. Individual galaxies are shown with lighter-colored points, while the data are divided into five bins and the median $x$ and $y$ values of each bin are shown with darker symbols to highlight the underlying trends. Black dashed lines indicate the best-fit linear relations. The Pearson correlation coefficient, $r$, and corresponding $p$ value are also included in each panel. The light blue line symbolizes the border at which the \NNI/\CaII and dust detections extend to the same radial distance from the center, and the shaded region indicates the region where the gas extends beyond dust detections. Only $\sim 20\%$ of \NaID detections extend beyond the dust, but for \CCII, this fraction jumps to $\sim 70\%$, suggesting that while \NaID depends on dust for shielding it against ionization, \CaII shows no such constraint and can easily extend beyond the limits set by dust.}
    \label{fig:NaD and CaII vs dust correlation radial distance}
\end{figure*}

Using this data, Figure \ref{fig:NaD and CaII vs dust correlation radial distance} shows that \NaID exhibits a weak but statistically significant dependence ($r=0.38$, $p=0.0044$) on dust; however, this correlation vanishes for \CaII ($r=-0.03$, $p=0.82$). Most interestingly, we find that while around $20\%$ of the galaxies host \NaID beyond the farthest dust measurements, for \CCII, this fraction jumps up to $\sim 70\%$! These results suggest that while \NaID depends on dust for shielding it against ionization, \CCII, with its higher ionization potential, does not indicate such a dependence because it is not as easily photoionized. We also note that in dusty regions, gas-phase Ca is more likely to be depleted onto dust grains, so some of the different distributions of dust and \CaII could be due to this.

\section{\textbf{Discussion}}
\label{section:discussion}

\begin{figure*}[t]
    \centering
    \includegraphics[width=0.98\textwidth]{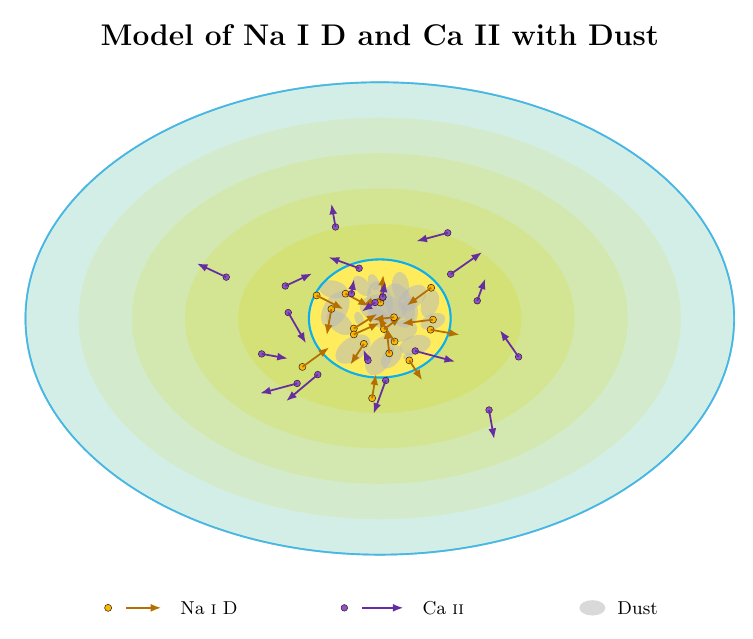} 
    \caption{Simple model of the \NaID and \CaII gas, in addition to dust, in our sample of quiescent galaxies. The shaded yellow regions in the background represent the stellar light profile of the galaxy, and the patchy grey ellipses represent dust. The markers represent \NaID (gold) and \CaII (purple) detections, while their arrows provide a general sense of their inflow/outflow kinematics and dispersion. \NaID tends to be more inflowing, coherent, and concentrated in the center, where the dust is mostly found. On the other hand, \CaII shows more outflows, much higher dispersion, and larger radial distributions. The blue shaded color corresponds to regions beyond where we detect dust. While \NaID is limited to where there is dust, \CaII shows no such dependence.}
    \label{fig:model of NaD and Ca II with dust}
\end{figure*}

\NaID ($\lambda\lambda 5891.58, 5897.56$ \AA) has an extensive history in being used to trace cool gas in galaxies, dating back to early surveys of super winds in local starburst and infrared-luminous galaxies \citep[e.g.,][]{2000ApJS..129..493H, 2005ApJ...631L..37R, 2005ApJS..160..115R, 2005ApJ...621..227M}. The use of \NaID in these systems was largely motivated by the strength of the \NaID absorption in a relatively clean part of the spectrum and helped establish that outflowing neutral gas is nearly ubiquitous in star-forming galaxies with high infrared luminosities. On the other hand, the \CaII H\&K, which was another tracer of cool interstellar gas was remained behind because it was located in a noisier, bluer part of the spectrum.

In recent years, \NaID has been used to trace not only outflows but also \textit{inflows} of neutral gas in quiescent systems (e.g., \citealt{2009ApJ...696..214S}). \citet{2021ApJ...919..145R} used \NaID to provide the first evidence for inflowing cool neutral gas in red geysers, a class of massive quiescent galaxies with weak, large-scale, AGN-driven ionized outflows \citep{2016Natur.533..504C, Roy_2018, 2021ApJ...913...33R}. This neutral gas was shown to be kinematically distinct from the ionized outflows (velocities of $\sim 40 \text{ km s}^{-1}$ vs. $\sim 300 \text{ km s}^{-1}$) and even the stars in their host galaxies ($W50_{NaD}/W50_* \sim 0.4$; \citealt{2021ApJ...913...33R, 2021ApJ...919..145R, 2026ApJ..1003...31M}). In Paper I, we further showed that the inflowing gas in red geysers is being used as a fuel supply for the AGN activity in the center, and that their presence was enhanced through interactions with nearby galaxies. In this work, we extend the analysis of Paper I by comparing the \NaID properties to those of the \CaII H\&K doublet in red geysers and a control sample of galaxies. 

Our discussion of the results is organized around three key observations: 1) detection prevalence, 2) velocity dispersion and velocity offset, and 3) connection between the ISM component and dust.

\textbf{Prevalence.} We find that \CaII is detected in substantially larger reservoirs than \NNI. This is both evident in our estimates of the areas of these reservoirs, the distributions of the radial extent of the gas from the center of the hosts, and the fraction of galaxies with detected \NaID vs. \CCII. This is consistent with absorption-line studies using background quasars, which tend to find more \CaII absorption in the outskirts than \NaID (e.g., \citealt{1991MNRAS.249..145B}, \citealt{2007MNRAS.379..738H}, \citealt{2011AJ....142..122C}, and \citealt{2022ApJ...936..171R}). These findings suggest that with \CCII, we are picking up a more common, less restrictive gas component than with \NNI, a point we return to when discussing dust.

The \NaID reservoirs that we detect in this work are slightly smaller than what we found in Paper I: we detect a total of 2643 spaxels with \NNI, roughly $90\%$ of the 2956 spaxels we found previously. This is likely because we are using different stellar templates and implementing different quality cuts when removing detections with low S/N. However, since our comparisons of \NaID vs. \CaII and red geysers vs. control galaxies in this work are conducted under the same templates and quality checks, we do not expect this systematic offset to affect our conclusions.

\textbf{Velocity Dispersion and Velocity Offset.} We find that \CaII traces gas with systematically larger dispersion than \NNI, while both have velocity dispersions lower than the stars.  Thus, \NaID seems to trace a more settled and dynamically colder gas component than \CCII. Our measured \NaID dispersion ($W50_{NaD}/W50_{*}\sim 0.35$) is slightly lower than what we measured in Paper I ($W50_{NaD}/W50_{*}\sim 0.44$), and our median velocity offset of $\sim 16.4 \text{ km s}^{-1}$ is slightly lower than what we estimated previously ($\sim 26.4 \text{ km s}^{-1}$). Similar to the prevalence measurements above, these small differences likely arise from the different quality cuts. Additionally, these distributions are broad. For instance, the velocity offset measurements vary by more than $200 \text{ km s}^{-1}$, and a $\sim 10 \text{ km s}^{-1}$ difference is minor and well within the range of uncertainties reported across the two measurements, as described in Section \ref{subsection:uncertainties and quality checks}. Indeed, when we compare the maps of \NaID velocities in red geysers in this work and in Paper I, the two are nearly identical in both their shape and their distribution of velocities. Our investigations show that the lower number of detections in this study is largely attributed to the requirement that $EW_{ISM}/EW_{tot}>0.056$, a conservative cut to avoid mistaking residuals from imperfect stellar template fits for genuine absorption from the ISM. Within these limits, our analysis shows that \NaID and \CaII are tracing kinematically distinct gas components in their host galaxies, with \CaII typically having dynamically warmer gas that is comparatively more outflowing than \NNI. It is intriguing that the kinematic differences between \NaID and \CaII are qualitatively similar to what is seen in the disk and halo of the Milky Way (\citet{1952ApJ...115..227R}), which has been attributed to the effects of shocks that liberate Ca from dust grains into the gas phase. 

\textbf{Dust.} We find a direct correlation between the size of the \NaID reservoir and that of the dust reservoir, a correlation that vanishes for \CCII. Similarly, we show that \NaID is typically confined to galactic radii at which we detect dust, but, once again, \CaII shows no such constraint, extending well beyond the dusty regions.

The difference in the dependence of \NaID and \CaII on dust provides deep physical insights about these gas phases. The ionization potential of the neutral sodium atom is only 5.14 eV, which means that it can easily be ionized in the presence of ultraviolet radiation. As a result, dust provides an effective tool for shielding and protecting the Na I. With the neutral sodium atom protected and present in these regions, we can detect the \NaID feature as it absorbs its background light. However, if this sodium atom goes beyond regions where there is no dust, it is easily ionized and destroyed, meaning we can no longer detect the \NaID feature that would arise from a neutral sodium atom. While this dust dependence limits where \NaID can be detected, it also makes the line a sensitive probe of dense, dusty gas specifically. \CCII, on the other hand, has a much higher ionization potential of 11.87 eV, allowing it to remain abundant (and detectable) even in regions where \NaID is easily ionized. This distinction helps explain why \CaII is much more prevalent than \NaID throughout our sample, and why the gas it traces can extend more easily to the outskirts of galaxies, unconstrained by the presence of dust, while \NaID remains confined to the dusty regions in the center. We also note that some of the differences between the associations of \NaID and \CaII with dust could arise because gas-phase Ca can be more strongly depleted onto dust than Na. 

We note two caveats to our dust measurements. First, since we rely on Balmer emission, our dust estimates are likely underestimated in these quiescent galaxies, where the emission is intrinsically faint. Second, our assumption of Case B recombination to convert the Balmer decrement into a dust estimate, may be imperfect if a fraction of the emission arises from shocked rather than purely photoionized gas, which could plausibly occur in red geysers with their bisymmetric outflows. However, shock models predict an intrinsic $H\alpha /H\beta$ ratio of $\sim 2.9-3.2$ \citep{2008ApJS..178...20A}, which is only slightly higher than our assumption of 2.86. Moreover, such shock-excited emission is expected to be confined to the bisymmetric outflow regions themselves rather than being distributed throughout the galaxy. With these considerations, we do not expect either of these to affect our qualitative conclusions regarding the differing dust dependence of \NaID and \CaII in our sample of galaxies.

All these results show that \NaID and \CaII cannot be treated as interchangeable tracers of gas in the ISM of galaxies. Their different sensitivities to dust, different spatial distributions, and different kinematics indicate that they are probing distinct gas phases with different physical origins. With their unique characteristics and distinctions, combining the two tracers provides an opportunity to gain a more complete picture of gas flows in galaxies, each with its own strengths: \NaID offers a relatively clean probe of the dense, dust-shielded neutral gas concentrated in the centers of galaxies, where it appears to trace ordered, settled gas with more inflows. In contrast, \CaII extends this picture to the more diffuse and dynamically warmer gas in the outskirts of galaxies, capturing material that \NaID would miss entirely. Neither tracer by itself can fully capture the complexity of the kinematics and distribution of the gases in galaxies; however, if used together, they can offer a more complete view of how cool gas is distributed, moves, and cycles through galaxies.

\section{\textbf{Summary}}
\label{sec: Summary}
In this work, we use spatially resolved data from SDSS MaNGA to study cool and warm gases traced by \NaID ($\lambda\lambda 5891.58, 5897.56$ \AA) and \CaII H\&K ($\lambda\lambda 3934.78, 3969.59$ \AA) in red geysers (local quiescent, active galaxies) and a control sample of galaxies. After removing the stellar absorption contributions, we use Gaussian fits to extract velocity and velocity dispersion of the gases and compare the two tracers of ISM gas, while also comparing red geysers to their matched control sample. We summarize our key findings below:
\begin{enumerate}
    \item \textbf{\CaII reservoirs are much larger than \NNI.} Our analysis of the size of the gas reservoirs shows that the \CaII reservoirs cover $\sim 2.5$ times larger areas than \NaID in red geysers. Similarly, in the control sample, the \CaII reservoirs are $\sim 6$ times larger than their \NaID counterparts. Overall, red geysers host more gas compared to the control sample. When comparing the control to the red geysers, we see a $\sim 4\times$ increase in reservoir size of \NaID and a $\sim 1.7\times$ increase in reservoir size of \CCII. 
    
    \item \textbf{\CaII traces gas with higher velocity dispersions.} When comparing the line width measurements of \NaID to \CCII, we find that the \CaII lines are $\sim 1.8$ times broader than \NaID in red geysers and $\sim 1.7$ times broader than \NaID in the control sample. We also find that the red geysers tend to have narrower absorption features at a statistically significant level, with \NaID in red geysers being $\sim 85\%$ as broad as \NaID in controls, while \CaII in red geysers is measured to be $\sim 93\%$ as broad as \CaII in controls. Both the \NaID and \CaII lines are systematically narrower than the stellar absorption lines. This has implications for the structure and origin of the gas.

    \item \textbf{Red geysers host larger dust reservoirs compared to the control galaxies.} We find that the dust-covered regions in red geysers are $\sim 2$ times larger in red geysers than the control sample (mean $Area/(\pi R_{50}^2)$: 0.174 vs. 0.082).

    \item \textbf{The extent/reservoir size of \NaID is linked to dust, whereas \CaII is not.} The size of the \NaID reservoir shows a modest but significant correlation with dust reservoir size ($r=0.35$, $p=4.3\times10^{-4}$), while the \CaII reservoir shows no significant correlation ($r=0.10$, $p=0.25$). Additionally, while in only $\sim 20\%$ of the galaxies there is \NaID beyond dust, this fraction jumps to $\sim 70\%$ for \CCII, suggesting that dust preferentially supports the survival of \NNI, which has a lower ionization potential (5.14 eV) than \CaII (11.87 eV).

\end{enumerate}

\appendix

\section{\textbf{Removal of H$\epsilon$ Contamination from the \CaII H Line}}
\label{appendix:hepsilon}
 
\subsection{Motivation}
\label{appendix:hepsilon:motivation}
Throughout our analysis, we restrict our study of the \CaII absorption to the K line at $3934.8\text{ \AA}$. This is because the \CaII H absorption line ($\lambda_{rest}=3969.6\text{ \AA}$) is only $\sim1.6 \text{ \AA}$ ($\sim$ 120 km/s) away from the H$\epsilon$ Balmer emission line ($\lambda_{rest}=3971.2\text{ \AA}$). This is close enough that the two will be blended together. As a result, any measurement of the H line would likely be weakened and distorted by H$\epsilon$ emission filling in the absorption. To address this contamination from H$\epsilon$, we use other Balmer lines (H$\alpha$ and H$\beta$) to predict and subtract the underlying H$\epsilon$ emission. We show that we can recover a physically plausible \CaII H line, but the correction is not robust enough to fully support any quantitative use of the H line. Still, this is a promising avenue for improving contamination from the H line, especially in higher S/N (or higher resolution) observations, and helps us explain why the H line appears so weak in the uncorrected spectra.

\subsection{Methodology}
\label{appendix:hepsilon:methodology}
To reconstruct and remove the H$\epsilon$, we proceed with the following steps:
\begin{enumerate}
    \item \textbf{H$\epsilon$ velocity.} Since the H$\alpha$ and H$\epsilon$ Balmer emission lines arise from the same excited hydrogen atoms, we assume that their velocities are equivalent, such that $v_{H\epsilon}=v_{H\alpha}$.

    \item \textbf{H$\epsilon$ velocity dispersion.} 
    We first measure the H$\alpha$ velocity dispersion and use the corresponding instrumental broadening to convert it to an intrinsic dispersion, 
    \begin{equation}
        \sigma_{H\alpha, intrinsic}=\sqrt{\sigma_{H\alpha, obs}^2 - \sigma_{H\alpha, inst}^2},
    \end{equation}
    which we take to be equal to the H$\epsilon$ dispersion ($\sigma_{H\epsilon, intrinsic} \equiv \sigma_{H\alpha, intrinsic}$). We then re-inject instrumental broadening at the wavelengths of H$\epsilon$ to get a predicted \textit{observed} dispersion,
    \begin{equation}
        \sigma_{H\epsilon, obs}=\sqrt{\sigma_{H\epsilon, intrinsic}^2 + \sigma_{H\epsilon, inst}^2}.
    \end{equation}

    \item \textbf{Observed H$\alpha$ and H$\beta$ fluxes.} Now that we have the velocity and dispersion of the H$\epsilon$, the only element left in re-constructing the Gaussian is the amplitude, which is related to the flux values of the H$\alpha$ and H$\beta$ thanks to atomic physics and is affected by dust reddening. We first take the observed H$\alpha$ and H$\beta$ fluxes by integrating over the observed emission features.
    
    \item \textbf{De-reddening to intrinsic flux.} Following \citet{2022ApJ...930..160S}, we convert our observed flux values for H$\alpha$ and H$\beta$ to intrinsic (dust-free) flux using the following expression:
    \begin{equation}
    \label{equation:F and f reddening}
        F_{\lambda} = f_{\lambda} \times 10^{0.4~k(\lambda)~E(B-V)},
    \end{equation}
    where $E(B-V)$ is the reddening and $k(\lambda)$ is calculated from the \citet{1994ApJ...422..158O} reddening curve assuming the V-band ratio of total to selective extinction, $R_v$=3.1. For H$\alpha$, the $k$ value is taken to be 2.52, and for H$\beta$, it is 3.66.

    \item \textbf{Predicting the intrinsic H$\epsilon$ flux.} Having calculated the intrinsic H$\alpha$ and H$\beta$ fluxes in step 4, we apply Case B recombination ratios (\citet{2006agna.book.....O}; $T\approx10,000 \text{ K}$). We first verify that $F(H\alpha)/F(H\beta) \approx 2.86$ in our sample and then adopt
    \begin{equation}
        F(H\epsilon )= 0.158~F(H\beta)
    \end{equation}
    for calculating the intrinsic H$\epsilon$ flux.

    \item \textbf{Predicting the observed H$\epsilon$ flux.} With the intrinsic H$\epsilon$ flux in hand, we seek to predict its \textit{observed} flux by taking into account the effects of dust. We first calculate H$\epsilon$'s extinction coefficient $k$ using \citet{1994ApJ...422..158O} with the assumption of $R_v=3.1$, which gives $k(H\epsilon)\approx4.44$. Then, we rearrange Equation \ref{equation:F and f reddening} in step 4 for a reddened, observed flux:
    \begin{equation}
        f_{H\epsilon} = \frac{F_{H\epsilon}}{10^{0.4~k(H\epsilon)~E(B-V)}}.
    \end{equation}

    \item \textbf{Gaussian amplitude.} Assuming a Gaussian profile for the H$\epsilon$ emission, we can extract the amplitude using the observed flux, which is the integral of the profile:
    \begin{equation}
        A=\frac{f_{H\epsilon}}{\sigma_{H\epsilon, obs}\sqrt{2\pi}}.
    \end{equation}

    \item \textbf{H$\epsilon$ construction and removal.} Having extracted the velocity, dispersion, and amplitude of the Gaussian profile, we can proceed with re-constructing the H$\epsilon$ emission:
    \begin{equation}
        f_{\text{H}\epsilon}(\lambda) = A \exp\left[-\frac{1}{2}\left(\frac{\lambda - \mu_{\text{obs}}}{\sigma_{\text{H}\epsilon,\text{obs}}}\right)^2\right],
    \end{equation}
    where $\mu_{obs}$ is the rest wavelength of H$\epsilon$ ($3971.2\text{ \AA}$) shifted by $v_{H\epsilon}$. We then subtract this profile from our observed spectrum, and finally divide the residual by the stellar continuum model from the DAP.

    \item \textbf{Line fitting.} Using the corrected spectrum, we fit the \CaII H\&K absorption features with a double-Gaussian model. Because the noise around the H line often makes the fitting routine more challenging, we first fit a single-Gaussian to the K line and use its best-fit parameters to put constraints on the double-Gaussian fits: 
    \begin{itemize}
        \item The amplitude of the blue K line is limited to be between 0.7 and 1.3 times its single-Gaussian value.
        
        \item The amplitude of the red H line is constrained to 0.5 to 0.99 times the blue-line amplitude (because of the ratio of their oscillator strength).
       
        \item The $W50$ is constrained to 0.7 to 1.3 times the single-Gaussian value.

        \item The velocity is constrained to within $\pm 70 \text{ km s}^{-1}$ of the single-Gaussian value.
    \end{itemize}
\end{enumerate}

\subsection{Recovering a Physically Plausible \CaII H Line and Its Limitations}
\label{appendix:hepsilon:results}
\begin{figure}[t]
    \centering
    \includegraphics[width=0.98\textwidth]{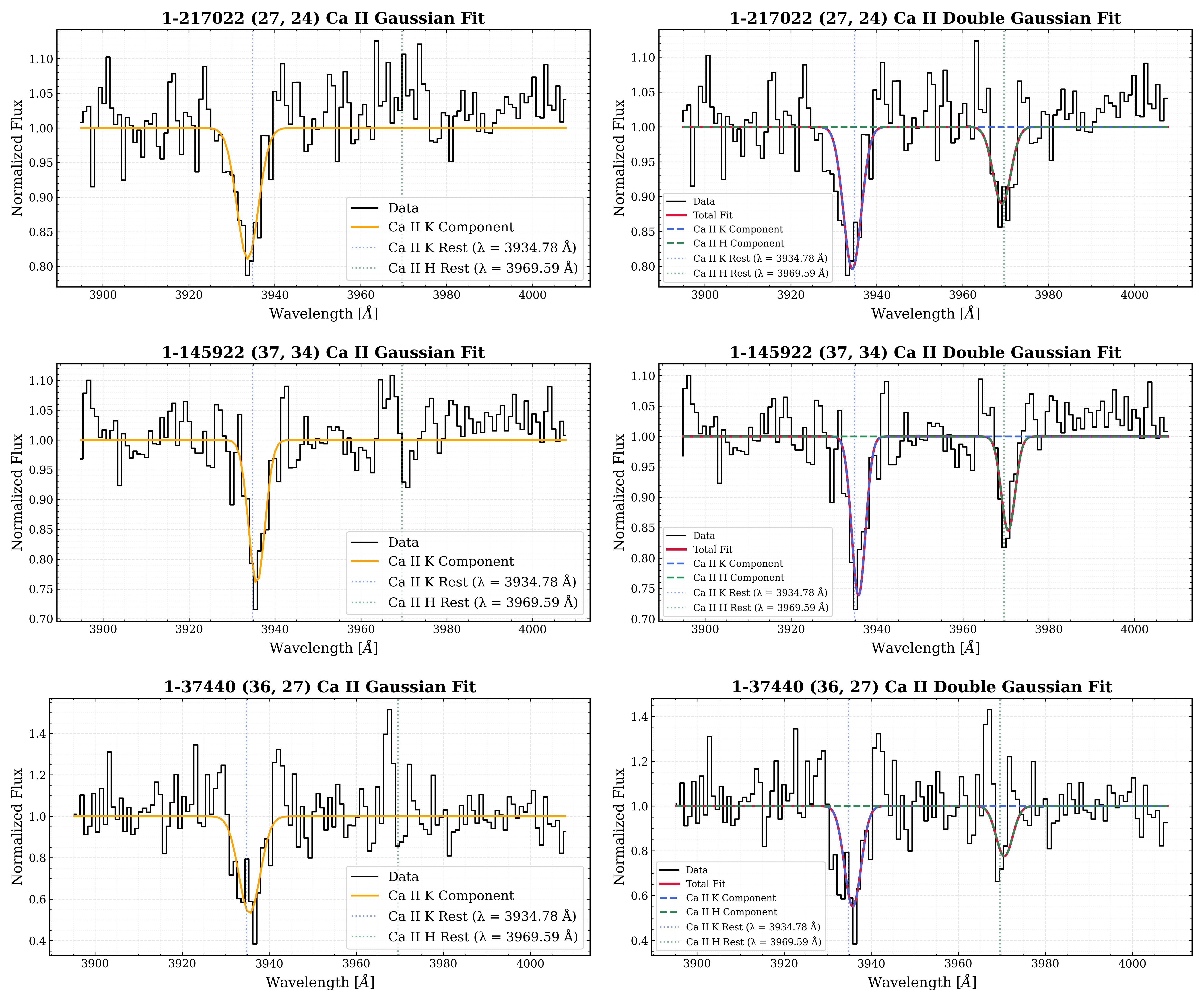} 
    \caption{Examples of single-Gaussian and double-Gaussian fits to the observed (left column) and H$\epsilon$-corrected (right column) \CaII spectra in three different spaxels across three rows:  galaxy 1-217022 at spaxel (27,24) in the top, galaxy 1-145922 at spaxel (37,34) in the middle, and galaxy 1-37440 at spaxel (36,27) in the bottom. In the left panels, we only fit a single Gaussian (orange) to the \CaII K line on the blue side. The corresponding rest wavelengths of the K-line ($3934.79 \text{ \AA}$) and H-line ($3969.59\text{ \AA}$) are shown with dotted vertical lines in blue and green, respectively. In the panels on the right-hand side, a double-Gaussian model (red) is fit to both the \CaII K (blue) and \CaII H (green) components. The double-Gaussian fits are constrained using single-Gaussian fits to the K line for better results, as described in Section \ref{appendix:hepsilon:methodology}. The observed (left) and cleaned (right) spectra show that the \CaII H line is contaminated by H$\epsilon$ emission. It is possible to remove the contamination from H$\epsilon$, but, as seen in the last row (an example of a bad fit), the signal-to-noise for the H line is still not high enough for robust measurements.}
    \label{fig:before and after H-epsilon fits}
\end{figure}

Three examples of our double-Gaussian fits to the cleaned \CaII H\&K are shown in Figure \ref{fig:before and after H-epsilon fits}. Before our corrections (in the left column), the \CaII H line at $\sim 3970 \text{ \AA}$ seems to be filled in by the H$\epsilon$ emission. However, after we construct and subtract the H$\epsilon$ emission at $\sim 3971\text{ \AA}$, a clear second trough appears that corresponds to a \CaII H component, with a depth and width comparable to the K line, consistent with the expected similarity of the two \CaII lines since they arise from a doublet. This indicates that the correction yields a physically plausible line profile.

Despite these results, we do not yet consider the corrected \CaII H line robust enough for any quantitative measurement (e.g., covering
fractions, optical depths, and equivalent widths). This is because in low S/N spaxels, the H-line region does not contain enough signal to be robustly used for double-Gaussian fits, even after our procedure for H$\epsilon$ correction.

As a result, we treat this exercise as a sanity check, showing that our overall methodology has the potential to recover a physically plausible \CaII H line and can explain why the H line appears so weak in the observed spectra before any corrections. All of our quantitative measurements and analyses on \CaII in this work are thus solely based on the uncontaminated \CaII K line.

\twocolumngrid
\begin{acknowledgments}
A.M. gratefully acknowledges the support of the Rowland Summer Research Fellowship at Johns Hopkins University and the Vivien Thomas Scholars Initiative for providing funding and making the research possible during the summers of 2024 and 2025.

Funding for the Sloan Digital Sky 
Survey IV has been provided by the 
Alfred P. Sloan Foundation, the U.S. 
Department of Energy Office of 
Science, and the Participating 
Institutions. 

SDSS-IV acknowledges support and 
resources from the Center for High 
Performance Computing  at the 
University of Utah. The SDSS 
website is www.sdss4.org.

SDSS-IV is managed by the 
Astrophysical Research Consortium 
for the Participating Institutions 
of the SDSS Collaboration including 
the Brazilian Participation Group, 
the Carnegie Institution for Science, 
Carnegie Mellon University, Center for 
Astrophysics | Harvard \& 
Smithsonian, the Chilean Participation 
Group, the French Participation Group, 
Instituto de Astrof\'isica de 
Canarias, The Johns Hopkins University, Kavli Institute for the 
Physics and Mathematics of the 
Universe (IPMU) / University of 
Tokyo, the Korean Participation Group, 
Lawrence Berkeley National Laboratory, 
Leibniz Institut f\"ur Astrophysik 
Potsdam (AIP),  Max-Planck-Institut 
f\"ur Astronomie (MPIA Heidelberg), 
Max-Planck-Institut f\"ur 
Astrophysik (MPA Garching), 
Max-Planck-Institut f\"ur 
Extraterrestrische Physik (MPE), 
National Astronomical Observatories of 
China, New Mexico State University, 
New York University, University of 
Notre Dame, Observat\'ario 
Nacional / MCTI, The Ohio State 
University, Pennsylvania State 
University, Shanghai 
Astronomical Observatory, United 
Kingdom Participation Group, 
Universidad Nacional Aut\'onoma 
de M\'exico, University of Arizona, 
University of Colorado Boulder, 
University of Oxford, University of 
Portsmouth, University of Utah, 
University of Virginia, University 
of Washington, University of 
Wisconsin, Vanderbilt University, 
and Yale University.

The Legacy Surveys consist of three individual and complementary projects: the Dark Energy Camera Legacy Survey (DECaLS; Proposal ID \#2014B-0404; PIs: David Schlegel and Arjun Dey), the Beijing-Arizona Sky Survey (BASS; NOAO Prop. ID \#2015A-0801; PIs: Zhou Xu and Xiaohui Fan), and the Mayall z-band Legacy Survey (MzLS; Prop. ID \#2016A-0453; PI: Arjun Dey). DECaLS, BAS,S and MzLS together include data obtained, respectively, at the Blanco telescope, Cerro Tololo Inter-American Observatory, NSF’s NOIRLab; the Bok telescope, Steward Observatory, University of Arizona; and the Mayall telescope, Kitt Peak National Observatory, NOIRLab. Pipeline processing and analyses of the data were supported by NOIRLab and the Lawrence Berkeley National Laboratory (LBNL). The Legacy Surveys project is honored to be permitted to conduct astronomical research on Iolkam Du’ag (Kitt Peak), a mountain with particular significance to the Tohono O’odham Nation.

NOIRLab is operated by the Association of Universities for Research in Astronomy (AURA) under a cooperative agreement with the National Science Foundation. LBNL is managed by the Regents of the University of California under contract to the U.S. Department of Energy.

This work also uses data obtained with the Dark Energy Camera (DECam), which was constructed by the Dark Energy Survey (DES) collaboration. Funding for the DES Projects has been provided by the U.S. Department of Energy, the U.S. National Science Foundation, the Ministry of Science and Education of Spain, the Science and Technology Facilities Council of the United Kingdom, the Higher Education Funding Council for England, the National Center for Supercomputing Applications at the University of Illinois at Urbana-Champaign, the Kavli Institute of Cosmological Physics at the University of Chicago, Center for Cosmology and Astro-Particle Physics at the Ohio State University, the Mitchell Institute for Fundamental Physics and Astronomy at Texas A\&M University, Financiadora de Estudos e Projetos, Fundacao Carlos Chagas Filho de Amparo, Financiadora de Estudos e Projetos, Fundacao Carlos Chagas Filho de Amparo a Pesquisa do Estado do Rio de Janeiro, Conselho Nacional de Desenvolvimento Cientifico e Tecnologico and the Ministerio da Ciencia, Tecnologia e Inovacao, the Deutsche Forschungsgemeinschaft and the Collaborating Institutions in the Dark Energy Survey. The Collaborating Institutions are Argonne National Laboratory, the University of California at Santa Cruz, the University of Cambridge, Centro de Investigaciones Energeticas, Medioambientales y Tecnologicas-Madrid, the University of Chicago, University College London, the DES-Brazil Consortium, the University of Edinburgh, the Eidgenossische Technische Hochschule (ETH) Zurich, Fermi National Accelerator Laboratory, the University of Illinois at Urbana-Champaign, the Institut de Ciencies de l’Espai (IEEC/CSIC), the Institut de Fisica d’Altes Energies, Lawrence Berkeley National Laboratory, the Ludwig Maximilians Universitat Munchen and the associated Excellence Cluster Universe, the University of Michigan, NSF’s NOIRLab, the University of Nottingham, the Ohio State University, the University of Pennsylvania, the University of Portsmouth, SLAC National Accelerator Laboratory, Stanford University, the University of Sussex, and Texas A\&M University.

BASS is a key project of the Telescope Access Program (TAP), which has been funded by the National Astronomical Observatories of China, the Chinese Academy of Sciences (the Strategic Priority Research Program “The Emergence of Cosmological Structures” Grant \# XDB09000000), and the Special Fund for Astronomy from the Ministry of Finance. The BASS is also supported by the External Cooperation Program of Chinese Academy of Sciences (Grant \# 114A11KYSB20160057), andthe  Chinese National Natural Science Foundation (Grant \# 12120101003, \# 11433005).

The Legacy Survey team makes use of data products from the Near-Earth Object Wide-field Infrared Survey Explorer (NEOWISE), which is a project of the Jet Propulsion Laboratory/California Institute of Technology. NEOWISE is funded by the National Aeronautics and Space Administration.

The Legacy Surveys imaging of the DESI footprint is supported by the Director, Office of Science, Office of High Energy Physics of the U.S. Department of Energy under Contract No. DE-AC02-05CH1123, by the National Energy Research Scientific Computing Center, a DOE Office of Science User Facility under the same contract; and by the U.S. National Science Foundation, Division of Astronomical Sciences under Contract No. AST-0950945 to NOAO.

\end{acknowledgments}

\bibliography{citations}{}
\bibliographystyle{aasjournalv7}

\end{document}